\documentclass[12pt,english,floatfix,superscriptaddress,aps,prd,preprint,nofootinbib]{revtex4}
\usepackage{amsmath}
\usepackage{amssymb}
\usepackage{amsbsy}
\usepackage[a4paper, margin=1.7cm]{geometry}
\usepackage{amsfonts}
\usepackage{amsopn}
\usepackage{amstext}
\usepackage{graphicx}
\usepackage{amssymb}
\usepackage{amsfonts}
\usepackage{amsmath}
\usepackage{graphicx}
\usepackage[english]{babel}
\usepackage{color}
\usepackage{xcolor}
\usepackage{slashed}
\usepackage{esint}
\usepackage[dvips]{epsfig}
\usepackage[dvips]{graphicx}
\usepackage{float}
\usepackage{units}
\usepackage{textcomp}
\usepackage{placeins}
\usepackage{hyperref}             
\hypersetup{
    colorlinks=true,              
    breaklinks=true,              
    citecolor=blue,               
    linkcolor=[rgb]{0,0.5,0.9},   
    urlcolor=red,                 
    filecolor=green               
}

\usepackage{hyperref}
\usepackage{slashed}

\newcommand{\ie}{\begin{equation}}
\newcommand{\fe}{\end{equation}}
 \newcommand{\bq}{\begin{equation}}
 \newcommand{\eq}{\end{equation}}
 \newcommand{\bqn}{\begin{eqnarray}}
 \newcommand{\eqn}{\end{eqnarray}}

\begin{document}

\title{Gravitational waves in \textit{metric--affine} bumblebee gravity}



\author{A. A. Ara\'{u}jo Filho}
\email{dilto@fisica.ufc.br}
\affiliation{Departamento de Física, Universidade Federal da Paraíba, Caixa Postal 5008, 58051--970, João Pessoa, Paraíba,  Brazil.}
\affiliation{Departamento de Física, Universidade Federal de Campina Grande Caixa Postal 10071, 58429-900 Campina Grande, Paraíba, Brazil.}
\affiliation{Center for Theoretical Physics, Khazar University, 41 Mehseti Street, Baku, AZ-1096, Azerbaijan.}

\date{\today}

\begin{abstract}

We study the propagation and emission of gravitational waves in the \textit{metric--affine} formulation of the bumblebee model, where spontaneous Lorentz symmetry breaking arises from a vector field acquiring a nonvanishing vacuum expectation value. Working in the geometric--optics limit of the linearized theory, we derive the modified dispersion relation governing the graviton modes and show that it depends on the orientation of the wave vector relative to the background vector. The polarization sector is examined for timelike and spacelike configurations of the Lorentz--violating vacuum. In both cases only two independent tensor modes propagate, although their propagation properties and tensor structure depend on the orientation of the background field. We then construct the retarded Green function associated with the modified wave operator and determine the radiation--zone produced by localized sources. In the timelike configuration the Lorentz--violating effects appear through a modified propagation speed and an overall amplitude renormalization, leading to a shifted retarded time while preserving the quadrupole structure of the waveform. In contrast, the spacelike sector introduces anisotropic corrections to the quadrupole amplitude together with an additional contribution proportional to the third time derivative of the quadrupole moment. As an astrophysical application, the gravitational radiation emitted by a circular binary black--hole system is evaluated, allowing observational constraints on the Lorentz--violating combination $\xi b^{2}$ to be estimated using multimessenger bounds from GW170817/GRB~170817A and waveform consistency requirements from gravitational wave observations.

\end{abstract}

\keywords{Gravitational waves; Lorentz symmetry breaking; polarization states; quadrupole term.}

\maketitle
\tableofcontents

\section{Introduction}

Investigations into the possible breakdown of Lorentz symmetry have attracted sustained attention in theoretical physics. One of the first motivations arose in string-motivated settings, where the dynamics of extended objects—particularly D-branes—may lead to a vacuum that selects preferred directions in spacetime, thereby generating spontaneous Lorentz violation in higher--dimensional constructions \cite{Kostelecky:1989jp,Kostelecky:1989jw,Kostelecky:1991ak,Kostelecky:1994rn,Gliozzi:2011hj}. Similar considerations later appeared in several quantum gravity proposals, such as loop quantum gravity scenarios, spacetime-foam descriptions, and models in which the propagation of fields is governed by modified dispersion relations \cite{Gambini:1998it,Calcagni:2016zqv,Bojowald:2004bb,Alfaro:2004aa,Amelino-Camelia:2001dbf}.

Because phenomena associated with the Planck scale cannot be probed directly with current experiments, a common approach is to describe possible residual effects through effective field theories that allow small departures from exact Lorentz invariance. Within this framework, the Standard Model Extension (SME) \cite{Colladay:1996iz,Colladay:1998fq} was formulated as a systematic construction containing Lorentz-- and CPT--violating operators built from the usual fields of particle physics. The formalism was later generalized to incorporate gravitational interactions in a low-energy relativistic setting \cite{Kostelecky:2003fs}. In many realizations, these terms arise when tensor fields develop nonvanishing vacuum expectation values, which introduce preferred directions in spacetime and thereby modify the symmetry structure at accessible energies.

In gravitational theories, imposing fixed background fields that define preferred spacetime directions can severely constrain the class of allowed solutions. Situations of this type appear, for instance, in frameworks such as Einstein–aether theory \cite{Jacobson:2000xp} and Chern Simons modified gravity \cite{Jackiw:2003pm}, where constant background structures may conflict with the dynamical nature of the geometry. A consistent alternative is obtained by treating the symmetry--breaking fields as dynamical variables. In that case the preferred directions emerge as solutions of the field equations, and diffeomorphism invariance remains intact because the background configuration is determined dynamically rather than imposed externally \cite{Jacobson:2000xp}.

Among the various realizations proposed for the gravitational sector of the SME, the bumblebee framework has received particular attention from the perspective of effective field theory \cite{Kostelecky:2003fs,Seifert:2009gi,Casana:2017jkc,Shi:2025rfq,Shi:2025plr,Ovgun:2018xys,AraujoFilho:2024iox,Amarilo:2023wpn,Maluf:2020kgf,Maluf:2014dpa}. In this construction the gravitational Lagrangian is built primarily from terms linear in the Ricci tensor, which represent the leading contributions at low energies. The model introduces a vector field governed by a potential that favors a nonvanishing vacuum configuration. Once such a state is reached, the vacuum selects a preferred spacetime direction and Lorentz symmetry is spontaneously broken. The vector field interacts with the gravitational sector through nonminimal couplings involving contractions with the Ricci tensor. 

In many studies of Lorentz--violating gravity, the geometric structure of spacetime is described within the standard metric formulation, where the metric tensor alone determines the properties of the gravitational field. A different perspective emerges in the \textit{metric--affine} (Palatini) approach, where the metric and the affine connection are regarded as independent dynamical quantities. 

Studies of Lorentz violation beyond the standard Riemannian framework have also considered geometries endowed with additional structures. Early work examined scenarios in which spacetime torsion or \textit{non--metricity} played a role in the breaking of Lorentz symmetry \cite{Kostelecky:2003fs,Foster:2016uui}. More general geometric settings, including Riemann–Finsler constructions, were later explored as possible arenas for such effects \cite{Kostelecky:2011qz}. In this broader context, a version of the bumblebee model formulated within the \textit{metric--affine} approach was proposed in \cite{Delhom:2019wcms}. The viability of the corresponding effective theory in the weak field regime—together with its couplings to scalar and Dirac matter—was subsequently analyzed and shown to be stable \cite{Delhom:2020gfv}. At the classical level, this formulation belongs to the wider family of Ricci based gravity theories, where matter sectors interact with the Ricci tensor through nonminimal couplings \cite{BeltranJimenez:2017doy,Delhom:2019zrb,Afonso:2018bpv,Afonso:2018mxn,BeltranJimenez:2019acz}.

More recently, \textit{non--metricity} has become an active ingredient in several developments related to black hole physics. Geometric frameworks where the connection is not metric-compatible have been used to construct new black hole solutions \cite{Filho:2022yrk,AraujoFilho:2024ykw}. Beyond the derivation of novel geometries, these setups have also provided a platform for examining thermodynamic behavior, optical signatures, and linear perturbations of compact objects in different scenarios \cite{Jha:2023vhn,AraujoFilho:2025hkm,Heidari:2024bvd,Shi:2025ywa,Lambiase:2023zeo,Filho:2024isd,Ovgun:2025mil}.

Gravitational waves are a fundamental prediction of general relativity, arising as propagating perturbations of spacetime generated by accelerated mass distributions. In the weak–field regime, their emission is described by the quadrupole formula, which relates the gravitational signal to the second time derivative of the source’s mass quadrupole moment \cite{Einstein:1916cc,Einstein:1918btx,Peters:1963ux}. The direct detection of gravitational waves from compact–binary mergers by the LIGO–Virgo–KAGRA network confirmed this prediction and established gravitational–wave observations as a powerful probe of strong–field gravity \cite{Abbott:2016blz,Abbott:2023lcc}. Because the waveform depends on the underlying gravitational dynamics, gravitational waves provide an important framework for testing modified theories of gravity \cite{Berti:2015itd,Yunes:2016jcc}.

They also offer a sensitive channel for investigating Lorentz symmetry in the gravitational sector. In effective frameworks such as the Standard–Model Extension, Lorentz–violating operators modify the dispersion relation of tensor perturbations and can produce anisotropic propagation, dispersion, or polarization effects in gravitational waves \cite{Kostelecky:2016kfm,Kostelecky:2018yfa,Mewes:2019dhj}. These signatures can be constrained using gravitational–wave observations, particularly through waveform consistency tests and multimessenger measurements such as GW170817, which impose strong limits on deviations of the gravitational wave speed from that of light \cite{Monitor:2017mdv,Abbott:2017oio}.

A recent study examined gravitational waves in the bumblebee gravity model within the \textit{metric} formulation \cite{Amarilo:2023wpn}, analyzing the polarization sector and anisotropic propagation. The corresponding analysis in the \textit{metric--affine} realization of the model remains unexplored. In this work we investigate gravitational wave propagation in the \textit{metric--affine} bumblebee scenario. Within the geometric--optics limit of the linearized theory, tensor perturbations obey a modified dispersion relation determined by the Lorentz--violating background vector. The polarization analysis shows that only two tensor modes propagate. In the timelike configuration the modification appears through a change in the propagation speed together with an overall rescaling of the waveform amplitude, while the spacelike sector introduces anisotropic corrections and an additional radiative term proportional to the third time derivative of the quadrupole moment.  The associated retarded Green function is constructed and used to determine the radiation--zone metric perturbation produced by localized sources. As an application, the gravitational radiation emitted by a circular binary black hole system is evaluated, allowing observational constraints on the parameter combination $\xi b^{2}$ to be inferred from multimessenger observations of GW170817/GRB~170817A and from waveform consistency in gravitational wave measurements.


\section{The \textit{metric--affine} bumblebee model}

The bumblebee model provides an effective field--theory realization of spontaneous Lorentz symmetry breaking. In this scenario a vector field $B_\mu$ is governed by a potential whose minimum occurs at $B_\mu B^\mu=\pm b^2$, so that the vacuum configuration $\langle B_\mu\rangle=b_\mu$ selects a preferred spacetime direction. The resulting nonvanishing vacuum expectation value breaks local Lorentz symmetry spontaneously while preserving general covariance.

In the \textit{metric--affine} formulation, the metric $g_{\mu\nu}$ and the affine connection $\Gamma^{\lambda}{}_{\mu\nu}$ are treated as independent variables, allowing \textit{non--metricity} and a more general geometric structure. In this framework the action of the \textit{metric--affine} bumblebee model can be written as
\begin{equation}
S_B =
\int \mathrm{d}^4x \sqrt{-g}
\left[
\frac{1}{2\kappa^2}
\Big(
R(\Gamma)
+
\xi B^\alpha B^\beta R_{\alpha\beta}(\Gamma)
\Big)
-\frac14 B_{\mu\nu}B^{\mu\nu}
-
V(B_\mu B^\mu \mp b^2)
\right]
+
S_m(g_{\mu\nu},\psi),
\label{actionMA}
\end{equation}
where $R(\Gamma)$ and $R_{\mu\nu}(\Gamma)$ denote the Ricci scalar and Ricci 
tensor constructed from the independent affine connection 
$\Gamma^{\lambda}{}_{\mu\nu}$, $\kappa^2=8\pi G$, and $B_{\mu\nu}=\partial_\mu B_\nu-\partial_\nu B_\mu$
is the field strength of the bumblebee vector field. The parameter $\xi$ 
controls the nonminimal coupling between the vector field and the Ricci tensor, 
while $V$ is the potential responsible for triggering spontaneous Lorentz 
symmetry breaking.

Independent variations of the action with respect to the metric, connection, 
and vector field yield the corresponding field equations
\begin{align}
\kappa^2 T_{\mu\nu}
&=
G_{(\mu\nu)}(\Gamma)
-\frac{\xi}{2}\,g_{\mu\nu}\,B^\alpha B^\beta R_{\alpha\beta}(\Gamma)
+2\xi\,B_{(\mu}R_{\nu)\beta}(\Gamma)B^\beta ,
\label{FE_metric}
\\[2mm]
0
&=
\nabla^{(\Gamma)}_{\lambda}
\!\left[
\sqrt{-g}\,
g^{\mu\alpha}
\left(
\delta^\nu_{\alpha}
+
\xi B^\nu B_\alpha
\right)
\right],
\label{FE_connection}
\\[2mm]
\nabla^{(g)}_{\mu}B^{\mu\nu}
&=
-\frac{\xi}{\kappa^2}\,
g^{\nu\alpha}B^\beta R_{\alpha\beta}(\Gamma)
+
2V'(B^2\mp b^2)B^\nu ,
\label{FE_B}
\end{align}
where $G_{(\mu\nu)}(\Gamma)$ denotes the symmetric part of the Einstein tensor 
constructed from the independent connection. The total energy--momentum tensor 
is given by $T_{\mu\nu}=T^M_{\mu\nu}+T^B_{\mu\nu}$, where $T^M_{\mu\nu}$ 
represents the contribution of the matter sources and
\begin{equation}
T^B_{\mu\nu}
=
B_{\mu\alpha}B_{\nu}{}^{\alpha}
-\frac14 g_{\mu\nu}B_{\alpha\beta}B^{\alpha\beta}
- V g_{\mu\nu}
+2V' B_\mu B_\nu
\end{equation}
is the contribution associated with the bumblebee field.

An important feature of the \textit{metric--affine} formulation emerges from the connection 
equation \eqref{FE_connection}. Since it contains no derivatives of the 
connection, this equation is algebraic and implies that the affine connection 
does not propagate independent degrees of freedom. It can therefore be solved 
explicitly, yielding the Levi--Civita connection of an auxiliary metric 
$h_{\mu\nu}$ defined through the disformal transformation
\begin{equation}
h_{\mu\nu}
=
\frac{1}{\sqrt{1+\xi X}}
\left(
g_{\mu\nu}
+
\xi B_\mu B_\nu
\right),
\qquad
X \equiv g^{\mu\nu}B_\mu B_\nu .
\label{disformal}
\end{equation}

Substituting this solution back into the action allows one to eliminate the 
connection and rewrite the theory in an Einstein--frame representation,
\begin{equation}
\widetilde S_{\rm BEF}
=
\int \mathrm{d}^4x \sqrt{-h}\,
\frac{1}{2\kappa^2}R(h)
+
S_m(h_{\mu\nu},B_\mu,\psi),
\label{EFaction}
\end{equation}
which corresponds to Einstein--Hilbert gravity for the auxiliary metric 
$h_{\mu\nu}$ coupled to a modified matter sector containing nonlinear 
interactions between the bumblebee field and the matter fields.

An important consequence of this representation is that the propagation of 
tensor perturbations is governed by the effective geometry defined by 
$h_{\mu\nu}$. In particular, gravitational waves propagate along the null cone 
of the Einstein--frame metric rather than the light cone of the original metric 
$g_{\mu\nu}$. The Lorentz--violating background encoded in the vector $b_\mu$ 
therefore induces a deformation of the effective causal structure experienced 
by gravitational perturbations.


\section{Linearized analysis and graviton dispersion relation}

\subsection{The linearized equations}

We start from the \textit{metric--affine} bumblebee theory written in its Einstein--frame (EF) representation, in which the independent connection has been integrated out and the gravitational sector reduces to the Einstein--Hilbert action for an auxiliary metric $h_{\mu\nu}$. The information about the original \textit{metric--affine} coupling is transferred to the matter sector. Concretely, the connection equation can be solved by introducing a disformally related metric
\begin{equation}
h_{\mu\nu}=\frac{1}{\sqrt{1+\xi X}}\Big(g_{\mu\nu}+\xi B_\mu B_\nu\Big),
\qquad
X\equiv g^{\mu\nu}B_\mu B_\nu,
\label{hdef}
\end{equation}
and the action can be recast as
\begin{equation}
\widetilde S_{\rm BEF}=\int \mathrm{d}^4x\,\sqrt{-h}\,\frac{1}{2\kappa^2}R(h)\;+\;{\widetilde S_m}(h_{\mu\nu},B_\mu,\psi),
\label{SEF}
\end{equation}
so that tensor perturbations propagate on the light cone defined by $h_{\mu\nu}$. In the weak--field regime this corresponds to $h_{\mu\nu}\simeq\eta_{\mu\nu}$ in the sense discussed in \cite{Delhom:2020gfv}.

In Ref.~\cite{Delhom:2020gfv} the weak--field analysis assumes $h_{\mu\nu}\approx\eta_{\mu\nu}$ and expands only the bumblebee field around a Lorentz--breaking vacuum. In order to analyze the propagation of tensor perturbations, however, the EF metric must remain dynamical. We therefore perform a simultaneous expansion of $(h_{\mu\nu},B_\mu)$ around Minkowski space and a constant vacuum expectation value,
$h_{\mu\nu}=\eta_{\mu\nu}+\kappa\,\widetilde h_{\mu\nu}$,
$B_\mu=b_\mu+\widetilde B_\mu$,
where $b_\mu$ is a constant background vector satisfying {$V'(X_0)=0$, with $X_0\equiv b^\mu b_\mu$}. The vector $b_\mu$ is treated as fixed background data, with indices raised and lowered by $\eta_{\mu\nu}$, so that $b^\mu=\eta^{\mu\nu}b_\nu$ and $b^2=\eta^{\mu\nu}b_\mu b_\nu$ are constants. We work to quadratic order in the fluctuations $(\widetilde h_{\mu\nu},\widetilde B_\mu)$, keeping the leading contributions in $\xi$ and the gravitational corrections in $\kappa$ relevant for the propagation of perturbations.

At the kinematical level the EF metric expansion gives
$h^{\mu\nu}=\eta^{\mu\nu}-\kappa\,\widetilde h^{\mu\nu}+\mathcal O(\kappa^2)$,
$\sqrt{-h}=1+\frac{\kappa}{2}\widetilde h+\mathcal O(\kappa^2)$,
with $\widetilde h\equiv\eta^{\mu\nu}\widetilde h_{\mu\nu}$. Following the EF weak--field structure of Ref.~\cite{Delhom:2020gfv}, the substitution $B_\mu=b_\mu+\widetilde B_\mu$ leads to a rescaled Maxwell--like kinetic term and an aether--type structure characterized by an effective metric $\widetilde\eta^{\mu\nu}$ and an effective mass tensor $M^{\mu}{}_{\nu}=-2\Lambda\,b^\mu b_\nu$. Taking the divergence yields the condition $(b\!\cdot\!\partial)(b\!\cdot\!\widetilde B)=0$, under which the free quadratic Lagrangian reduces to Eq.~(15) of \cite{Delhom:2020gfv}.

To generalize this structure with dynamical $h_{\mu\nu}$, we promote the contractions $\eta^{\mu\nu}\to h^{\mu\nu}$ to the required order while keeping $b_\mu$ fixed. We also define the fluctuation field strength
$\widetilde B_{\mu\nu}\equiv\partial_\mu\widetilde B_\nu-\partial_\nu\widetilde B_\mu$,
which involves only partial derivatives and therefore does not require introducing Christoffel symbols.

The EF quadratic Lagrangian density for $\widetilde B_\mu$ in the broken vacuum is obtained by taking the same EF weak--field structure of \cite{Delhom:2020gfv} but promoting explicit $\eta$--contractions to $h$--contractions and expanding to first order in $\kappa$. In particular, the effective tensor entering the kinetic operator becomes
$\widetilde\eta_{(h)}^{\mu\nu}
=
\Big(1+\frac{\xi}{2}b^2\Big)\,h^{\mu\nu}-2\xi\,b^\mu b^\nu$,
so that, expanding around Minkowski space,
$\widetilde\eta_{(h)}^{\mu\nu}
=\widetilde\eta^{\mu\nu}+\kappa\,\delta\widetilde\eta^{\mu\nu}+\mathcal O(\kappa^2)$,
$\widetilde\eta^{\mu\nu}
=
\Big(1+\frac{\xi}{2}b^2\Big)\eta^{\mu\nu}-2\xi\,b^\mu b^\nu$,
$\delta\widetilde\eta^{\mu\nu}
=
-\Big(1+\frac{\xi}{2}b^2\Big)\widetilde h^{\mu\nu}$.
This makes explicit how the {EF metric} fluctuation $\widetilde h_{\mu\nu}$ modifies the effective tensor structure already present in the EF weak--field {vector sector}.

At quadratic order in $\widetilde B_\mu$ (and to first order in $\xi$), the $h$--covariant generalization of the free EF vector Lagrangian in the broken vacuum can be written as
\begin{equation}
\mathcal L_{B}^{(2)}
=
-\frac{1}{4}\sqrt{-h}\,\widetilde\eta_{(h)}^{\mu\alpha}\,
\widetilde B_{\mu\nu}\,\widetilde B_{\alpha\rho}\,h^{\nu\rho}
-
\sqrt{-h}\,\Lambda\,\big(b^{\mu}\widetilde B_\mu\big)^2
+\mathcal O(\xi^2).
\label{LB2}
\end{equation}

Varying \eqref{LB2} with respect to $\widetilde B_\mu$ and keeping only terms up to $\mathcal O(\kappa)$ yields the generalization of the free vector equations. In the strict Minkowski limit ($\kappa\to0$) one recovers the operator quoted in \cite{Delhom:2020gfv},
\begin{equation}
0=\partial_\mu\widetilde B^{\mu\nu}\Big(1+\frac{\xi}{2}b^2\Big)
-\xi\,b^\mu b^\alpha\partial_\mu \widetilde B_{\alpha}{}^{\nu}
+\xi\,b^\nu b^\alpha\partial_\mu \widetilde B_{\alpha}{}^{\mu}
-2\Lambda\,b^\nu b^\alpha \widetilde B_\alpha,
\label{EOMflat}
\end{equation}
and its compact form in terms of the effective metric.

When $\kappa\neq0$, additional terms arise from the expansion of $\sqrt{-h}$, from the replacement $\widetilde\eta^{\mu\nu}\to\widetilde\eta_{(h)}^{\mu\nu}$, and from raising indices with $h^{\mu\nu}$. The resulting equations can be organized schematically as
$\mathcal E^\nu(\widetilde B;\eta,b)
+\kappa\,\delta\mathcal E^\nu(\widetilde B;\widetilde h,\eta,b)=0$,
where $\mathcal E^\nu(\widetilde B;\eta,b)=0$ corresponds to \eqref{EOMflat} and the correction $\delta\mathcal E^\nu$ is linear in $\widetilde h_{\mu\nu}$.

Taking the divergence of the deformed equation yields the modified constraint. In the strict Minkowski case \cite{Delhom:2020gfv} obtains
\begin{equation}
(b\cdot\partial)(b\cdot\widetilde B)=0.
\label{constraint0}
\end{equation}
With
$h_{\mu\nu}=\eta_{\mu\nu}+\kappa\widetilde h_{\mu\nu}$,
the corresponding covariantized statement becomes
$(b\cdot\partial)(b\cdot\widetilde B)=\mathcal O(\kappa)$.
To the order relevant for the propagation analysis we may therefore impose the zeroth--order constraint \eqref{constraint0}.

In this way the EF quadratic Lagrangian can be written as
\begin{align}
\mathcal L_{B}^{(2)}
&=
\mathcal L_{B,0}^{(2)}
+\kappa\,\delta\mathcal L_{B}^{(2)}
+\mathcal O(\kappa^2),
\label{Lsplit}
\\
\mathcal L_{B,0}^{(2)}
&=
-\frac14
\left(1+\frac{\xi}{2}b^2\right)
\widetilde B_{\mu\nu}\widetilde B^{\mu\nu}
+
\frac{\xi}{2}
b^\mu b^\alpha
\widetilde B_{\mu\nu}
\widetilde B_{\alpha}{}^{\nu}
-
\Lambda
(b^\alpha \widetilde B_\alpha)^2
+
\mathcal O(\xi^2),
\label{L0}
\\
\delta\mathcal L_{B}^{(2)}
&=
\frac12\,\widetilde h\,\mathcal L_{B,0}^{(2)}
-\frac14\,\delta\widetilde\eta^{\mu\alpha}\widetilde B_{\mu\nu}\widetilde B_{\alpha}{}^{\nu}
+\frac14\,\widetilde\eta^{\mu\alpha}\widetilde h^{\nu\rho}\widetilde B_{\mu\nu}\widetilde B_{\alpha\rho}
+\mathcal O(\xi^2),
\label{Ld_corr}
\end{align}
with
\begin{equation}
\widetilde\eta^{\mu\nu}
=
\Big(1+\frac{\xi}{2}b^2\Big)\eta^{\mu\nu}-2\xi\,b^\mu b^\nu,
\qquad
\delta\widetilde\eta^{\mu\nu}
=
-\Big(1+\frac{\xi}{2}b^2\Big)\widetilde h^{\mu\nu}.
\end{equation}
{The tensor $\widetilde\eta^{\mu\nu}$ characterizes the kinetic structure of the bumblebee fluctuation $\widetilde B_\mu$ and should not be identified with the effective tensor controlling the characteristic cone of the spin--2 perturbation.}

A complete derivation of the graviton propagator would require constructing the {full} quadratic wave operator for {the tensor fluctuation} $\widetilde h_{\mu\nu}$ and inverting it using {an enlarged} Barnes--Rivers spin projector {basis adapted to the constant background vector $b_\mu$}, as in Ref.~\cite{Maluf:2014dpa}. Because the background vector $b_\mu$ introduces additional tensor structures, the operator basis becomes enlarged and the inversion procedure becomes technically involved.

For the purposes of the present work such a full inversion is not required. {The goal is not to obtain the complete graviton propagator, but only the characteristic cone associated with tensor propagation.} {This cone} can be extracted from the principal part of the linearized {tensor} equations in the geometric--optics (eikonal) limit, where the perturbations are treated as rapidly oscillating waves. Retaining only the highest--derivative terms allows one to identify the characteristic equation and determine the modified dispersion relation for the graviton modes.

\subsection{The modified dispersion relation}

A practical way to determine the graviton dispersion relation without constructing the full propagator is to analyze the \emph{principal symbol} (or characteristic polynomial) of the linearized field equations. This approach focuses on the highest--derivative structure of the equations of motion and therefore directly identifies the characteristic hypersurfaces along which disturbances propagate. In field theory language, this corresponds to the geometric--optics (or eikonal) limit, in which the wavelength of the perturbation is assumed to be much smaller than the characteristic curvature scale of the background.

To implement this procedure, one considers a rapidly oscillating perturbation of the EF metric fluctuation,
$\widetilde h_{\mu\nu}(x)
=
\Re\!\left\{
\varepsilon_{\mu\nu}(x)\,e^{i\Theta(x)/\epsilon}
\right\},
\,
k_\mu \equiv \partial_\mu \Theta,
\,
\epsilon \to 0$,
where $\Theta(x)$ denotes the rapidly varying phase and $\varepsilon_{\mu\nu}(x)$ is a slowly varying polarization tensor. In the limit $\epsilon\to 0$, derivatives acting on the exponential dominate over derivatives acting on the amplitude, so that the leading contribution to the equations of motion arises from the highest--derivative terms. Operationally, this corresponds to replacing
$\partial_\mu \to i\,k_\mu$
in the principal part of the linearized operator. The resulting equations become algebraic in $k_\mu$, and the condition for the existence of nontrivial solutions is obtained from the vanishing of the determinant of the principal symbol, $\det\!\left[\sigma\!\left(\widehat{\mathcal O}_{hh}\right)(k)\right]=0$,
which defines the dispersion relation.

In the Einstein--frame representation, the gravitational sector of the theory is purely Einstein--Hilbert in terms of the auxiliary metric $h_{\mu\nu}$. Consequently, the highest--derivative part of the linearized spin--2 operator coincides with the standard Einstein--Hilbert kinetic structure evaluated on the EF background metric. Since the characteristic structure of the equations depends only on this principal part, the propagation of tensor perturbations is governed by the effective background geometry defined by $h_{\mu\nu}$.

In the Lorentz--violating vacuum of the bumblebee model, one takes
$g_{\mu\nu}=\eta_{\mu\nu},
\,
B_\mu=b_\mu=\text{const.}$,
so that the disformal relation \eqref{hdef} determines a constant EF background metric,
\begin{equation}
\bar h_{\mu\nu}
=
\frac{1}{\sqrt{1+\xi b^2}}
\Big(
\eta_{\mu\nu}
+
\xi\, b_\mu b_\nu
\Big),
\qquad
b^2 \equiv \eta^{\mu\nu} b_\mu b_\nu .
\end{equation}
Because this background is constant, the principal symbol of the linearized Einstein operator reduces to the standard wave operator constructed with $\bar h_{\mu\nu}$. Up to gauge terms---which remove unphysical polarizations but do not modify the characteristic cone---the propagation of the spin--2 perturbation is therefore determined by the null condition associated with the EF background metric. The graviton dispersion relation follows directly from
$\bar h^{\mu\nu} k_\mu k_\nu = 0$.

This result shows that the tensor perturbations propagate along the null cone defined by the EF metric rather than the Minkowski metric. In other words, the Lorentz--violating background encoded in the vector $b_\mu$ effectively deforms the propagation cone of the graviton.

Expanding the inverse EF metric to first order in the coupling parameter $\xi$ yields
\begin{equation}
\bar h^{\mu\nu}
=
\eta^{\mu\nu}
-
\xi\!\left(
b^\mu b^\nu
-
\frac12 b^2\,\eta^{\mu\nu}
\right)
+
\mathcal O(\xi^2),
\end{equation}
so that the null condition becomes, to leading order in $\xi$,
\begin{equation}
k^2
-
\xi\Big[
(b\!\cdot\!k)^2
-
\frac12 b^2 k^2
\Big]
=0 .
\label{disp_grav_eikonal}
\end{equation}
This expression represents the modified dispersion relation for the graviton in the Lorentz--violating vacuum of the \textit{metric--affine} bumblebee theory. It explicitly shows that the propagation depends on the orientation of the wave vector relative to the background vector $b_\mu$, leading to an anisotropic deformation of the effective light cone.

For completeness, we present in Tab.~\ref{tabbbb} a comparison between the graviton results obtained in the \textit{metric--affine} (this work) and purely \textit{metric} (Maluf et. at. \cite{Maluf:2014dpa}) formulations of the bumblebee model. In particular, the table summarizes the origin of the modifications, the role of the effective metric, and the corresponding dispersion relations in each approach.

\begin{table}[h]
\centering
\small
\begin{tabular}{||p{4.5cm}||p{5.2cm}||p{5.2cm}||}
\hline\hline\hline
 & \quad\quad\quad \textbf{\textit{Metric--Affine}  } & \quad\quad \quad\quad \textbf{\textit{Metric}} \cite{Maluf:2014dpa} \\
\hline\hline\hline

\,\textbf{Origin of modification}
&
Propagation on the null cone of the effective Einstein--frame metric $h_{\mu\nu}$.
&
Direct modification of the graviton kinetic operator due to the Lorentz--violating background vector $b_\mu$.
\\

\hline

\quad\quad\textbf{Effective metric}
&
\[
h_{\mu\nu}=
\frac{1}{\sqrt{1+\xi b^2}}
\left(\eta_{\mu\nu}+\xi b_\mu b_\nu\right)
\]
&
No effective metric appears. Lorentz violation modifies the wave operator directly.
\\

\hline

\quad\textbf{Dispersion relation}
&
\[
h^{\mu\nu}k_\mu k_\nu=0
\]
&
\[
k^2+\xi (b\!\cdot\!k)^2=0
\]

\\

\hline

\quad\quad\textbf{Expanded form}
&
\[
k^2-\xi\Big[(b\!\cdot\!k)^2-\frac12 b^2 k^2\Big]=0
\]
&
\[
k^2+\xi(b\!\cdot\!k)^2=0
\]
\\

\hline\hline\hline

\end{tabular}
\caption{Comparison between the modified graviton dispersion relations in the \textit{metric--affine} bumblebee theory (this work) and the \textit{metric} formulation studied by Maluf et. al. \cite{Maluf:2014dpa}\label{tabbbb}}
\end{table}

\section{The polarization states}

Following the same procedure adopted in Ref.~\cite{Amarilo:2023wpn}, we analyze the polarization of the free graviton by considering a plane--wave ansatz
\begin{equation}
{\widetilde h}_{\mu\nu}(x)=\varepsilon_{\mu\nu}\,e^{ik\cdot x},\qquad
k^\mu=(\omega,0,0,k),\qquad k\cdot x=\omega t-kz .
\end{equation}
The polarization tensor $\varepsilon_{\mu\nu}$ encodes the tensor degrees of freedom of the gravitational wave. In the \textit{metric} formulation of the bumblebee model~\cite{Amarilo:2023wpn}, the linearized field equations lead to a set of algebraic constraints that remove unphysical components of the tensor perturbation. In particular, those equations imply that no polarization is allowed along the Lorentz--violating background direction,
\begin{equation}
b^\mu {\widetilde h}_{\mu\nu}=0
\qquad\Longrightarrow\qquad
b^\mu \varepsilon_{\mu\nu}=0 ,
\label{bconstraint}
\end{equation}
together with the transversality condition $k^\mu \varepsilon_{\mu\nu}=0$. In addition, the trace condition \begin{equation} \varepsilon\equiv \eta^{\mu\nu}\varepsilon_{\mu\nu}=0 \label{tracecond} \end{equation} is imposed whenever it follows from the on--shell constraint system. For certain special configurations, however, the constraint structure becomes degenerate and the trace is not forced to vanish.

In the present \textit{metric--affine} bumblebee scenario, the dispersion relation is obtained from the principal symbol of the linearized equations in the geometric--optics limit. A complete determination of the polarization sector would require the explicit construction and inversion of the quadratic spin--2 operator, i.e.\ the full graviton propagator, in the presence of the background vector $b_\mu$, along the lines of the analysis performed in Ref.~\cite{Maluf:2014dpa}. Since the main goal of the present work is to determine the propagation law and its consequences for gravitational radiation, we restrict the discussion to representative polarization structures compatible with the transverse tensor sector. Accordingly, the explicit tensor forms written below should be viewed as illustrative representatives inspired by the \textit{metric} analysis, rather than as a complete derivation of the physical polarization content of the \textit{metric--affine} theory.

Within this approximation, the physical pole is governed by the modified dispersion relation
$k^2-\xi\left[(b\!\cdot\!k)^2-\frac12 b^2 k^2\right]=0$ ,
which can be written as
\begin{equation}
\Big(1+\frac{\xi}{2}b^2\Big)\,k^2-\xi\,(b\!\cdot\!k)^2=0 .
\label{MDR_MA_rewrite}
\end{equation}


\subsection{Timelike background: $b^\mu=(b_0,0,0,0)$}

Adopting the representative constraint structure $b^\mu\varepsilon_{\mu\nu}=0$ together with $k^\mu\varepsilon_{\mu\nu}=0$, one obtains $\varepsilon_{0\nu}=0$ and $\varepsilon_{3\nu}=0$, while Eq.~\eqref{tracecond} yields $\varepsilon_{11}+\varepsilon_{22}=0$. Therefore the polarization tensor reduces to the two standard transverse modes
\begin{equation}
\varepsilon_{\mu\nu}=
\begin{pmatrix}
0&0&0&0\\
0&\varepsilon_{11}&\varepsilon_{12}&0\\
0&\varepsilon_{12}&-\varepsilon_{11}&0\\
0&0&0&0
\end{pmatrix}.
\label{pol_timelike}
\end{equation}

The propagation is modified by the dispersion relation. Using $b\!\cdot\!k=b_0\,\omega$ and $b^2=b_0^2$, Eq.~\eqref{MDR_MA_rewrite} gives
\begin{equation}
\omega^2
=
k^2
\frac{1+\frac{\xi}{2}b_0^2}{1-\frac{\xi}{2}b_0^2},
\qquad
\omega
=
k\sqrt{\frac{1+\frac{\xi}{2}b_0^2}{1-\frac{\xi}{2}b_0^2}} .
\label{disp_timelike}
\end{equation}


\subsection{Spacelike background parallel to propagation: $b^\mu=(0,0,0,b_3)$}

In this configuration, adopting the same representative constraint structure, one obtains $\varepsilon_{3\nu}=0$. Together with the transversality and trace conditions, this again yields the same two transverse modes as in Eq.~\eqref{pol_timelike}. The dispersion relation now involves $b\!\cdot\!k=-b_3 k$ and $b^2=-b_3^2$, yielding
\begin{equation}
\omega^2
=
k^2
\frac{1+\frac{\xi}{2}b_3^2}{1-\frac{\xi}{2}b_3^2},
\qquad
\omega
=
k\sqrt{\frac{1+\frac{\xi}{2}b_3^2}{1-\frac{\xi}{2}b_3^2}} .
\label{disp_spacelike_parallel}
\end{equation}


\subsection{Spacelike background orthogonal to propagation: $b^\mu=(0,0,b_2,0)$}

For this configuration $b\!\cdot\!k=0$ and $b^2=-b_2^2$, so Eq.~\eqref{MDR_MA_rewrite} reduces to $\Big(1-\frac{\xi}{2}b_2^2\Big)(\omega^2-k^2)=0 \,\Rightarrow\, \omega=k$, provided $1-\xi b_2^2/2\neq0$. Hence the dispersion relation remains Lorentz invariant. Adopting again the representative constraint structure used in the metric analysis~\cite{Amarilo:2023wpn}, the condition $b^\mu\varepsilon_{\mu\nu}=0$ removes the entire $y$ row and column, $\varepsilon_{2\nu}=0$ . In this configuration the constraint system does not enforce $\varepsilon=0$, so the trace component may survive. Solving the remaining relations yields a two--parameter polarization tensor (e.g.\ with $\varepsilon_{11}$ and $\varepsilon_{13}$ independent),
\begin{equation}
\varepsilon_{\mu\nu}=
\begin{pmatrix}
-\frac12\,\varepsilon_{11} & -\varepsilon_{13} & 0 & \frac12\,\varepsilon_{11}\\
-\varepsilon_{13} & \varepsilon_{11} & 0 & \varepsilon_{13}\\
0 & 0 & 0 & 0\\
\frac12\,\varepsilon_{11} & \varepsilon_{13} & 0 & -\frac12\,\varepsilon_{11}
\end{pmatrix}.
\label{pol_spacelike_perp}
\end{equation}


In other words, {within the representative \textit{metric-}-inspired constraint structure adopted here, the illustrative tensor sector is parametrized by two independent amplitudes}. Nevertheless, their tensor representation depends on the orientation of the background vector relative to the propagation direction. For timelike backgrounds or spacelike backgrounds parallel to the wave vector, {the representative tensor structure reduces to the usual transverse form, while the propagation velocity is modified}. In contrast, when $b^\mu\perp k^\mu$ the dispersion relation becomes Lorentz invariant, while a representative tensor description inspired by the metric analysis may contain additional coordinate components. A complete identification of the physical polarization content in the metric--affine theory would require the full graviton propagator.


\subsection{Geometric and tidal interpretation for $b^\mu\perp k^\mu$}

In the orthogonal configuration $b^\mu=(0,0,b_2,0)$, one has $b\!\cdot\!k=0$ and Eq.~\eqref{MDR_MA_rewrite} implies the Lorentz--invariant dispersion relation $k^0=k^3$, i.e.\ $\omega=k$, in complete analogy with the metric result of Ref.~\cite{Amarilo:2023wpn}. At the illustrative level, and adopting the representative tensor structure \eqref{pol_spacelike_perp}, the plane wave ${\widetilde h}_{\mu\nu}=\varepsilon_{\mu\nu}e^{i(\omega t-kz)}$ inserted into the perturbed line element $\mathrm{d}s^2=(\eta_{\mu\nu}+{\widetilde h}_{\mu\nu})\,\mathrm{d}x^\mu \mathrm{d}x^\nu$ yields, to linear order in the amplitudes, {where $\varepsilon_{11}$ and $\varepsilon_{13}$ denote the corresponding real oscillatory perturbations, $\mathrm{Re}[\varepsilon_{11}e^{i(\omega t-kz)}]$ and $\mathrm{Re}[\varepsilon_{13}e^{i(\omega t-kz)}]$, respectively,}
\begin{align}
\mathrm{d}s^2
&=
\Big(1-\frac12\,\varepsilon_{11}\Big)\,\mathrm{d}t^2
-\Big(1-\varepsilon_{11}\Big)\,\mathrm{d}x^2
-\mathrm{d}y^2
-\Big(1+\frac12\,\varepsilon_{11}\Big)\,\mathrm{d}z^2
\nonumber\\
&\hspace{8mm}
+2\varepsilon_{13}\,\mathrm{d}x\,\mathrm{d}z
-2\varepsilon_{13}\,\mathrm{d}t\,\mathrm{d}x
+\varepsilon_{11}\,\mathrm{d}t\,\mathrm{d}z .
\label{ds2_MA_perp}
\end{align}

Then, the representative tensor structure produces contractions and dilatations in the $xz$ plane and also induces time deformations at the coordinate level. The effects on particles can be characterized through the geodesic deviation equation. For slowly moving test particles with $U^\mu\simeq(1,0,0,0)$, it reduces to
\begin{equation}
\frac{D^2 S^\mu}{\mathrm{d}\tau^2}=R^\mu{}_{00\sigma}\,S^\sigma,
\label{geo_dev_MA}
\end{equation}
where $S^\mu$ is the separation vector between neighboring geodesics. Here and in what follows we use ${\widetilde h}^{\mu}{}_\nu\equiv \eta^{\mu\alpha}{\widetilde h}_{\alpha\nu}$. Using the representative polarization tensor \eqref{pol_spacelike_perp}, the temporal and $y$ components obey $ \frac{D^2 S^0}{\mathrm{d}\tau^2}=0, \,\, \frac{D^2 S^2}{\mathrm{d}\tau^2}=0 $, so that no net tidal force arises along the $y$ direction, and no time--tilde acceleration appears in this illustrative setup (even though ${\widetilde h}_{00}\neq0$).

In the $x$ direction, Eq.~\eqref{geo_dev_MA} gives
\begin{equation}
\frac{D^2 S^1}{\mathrm{d}\tau^2}
=
{
-\frac{\omega^2}{2}\,
\widetilde h^{1}{}_{1}\,S^1
},
\label{tidal_x_MA}
\end{equation}
whereas the tidal force in the longitudinal ($z$) direction takes the form
${
\frac{D^2 S^3}{\mathrm{d}\tau^2}
=
0
}$.

These expressions should be interpreted as a coordinate--level illustration associated with the representative tensor \eqref{pol_spacelike_perp}. Again, as we commented before, a rigorous identification of the physical polarization content and its tidal effects in the \textit{metric--affine} theory would require the full graviton propagator. When the separation between two test particles is sufficiently small, their events may be treated as occurring simultaneously, allowing the timelike component $S^0$ to be disregarded. Under this approximation, {the representative tidal response in the $x$ direction is controlled by the $\varepsilon_{11}$ amplitude, while no longitudinal tidal displacement is produced at this level}. {The $\varepsilon_{13}$ component should therefore be interpreted only as part of the coordinate--level representative tensor structure, not as a separately established observable longitudinal mode in the \textit{metric--affine} theory.}

\section{Emission of gravitational waves}


\subsection{The timelike case}

Following the same strategy adopted for the construction of the retarded Green function in the minimal gravitational SME framework \cite{AraujoFilho:2026zyt,AraujoFilho:2026vcf}, we consider the modified dispersion relation arising in the \textit{metric--affine} bumblebee gravity scenario. In this case the propagation of tensor perturbations is governed by Eq. (\ref{disp_grav_eikonal}).  Let us first examine the timelike configuration of the background vector,
\begin{equation}
b^\mu=(b_0,0,0,0),
\end{equation}
so that
\begin{equation}
b^2=b_0^2,
\qquad
b\!\cdot\!k=b_0\,\omega .
\end{equation}
Substituting these relations into the dispersion relation yields
\begin{equation}
(\omega^2-\mathbf{k}^2)
-\xi\left[b_0^2\omega^2-\frac{1}{2}b_0^2(\omega^2-\mathbf{k}^2)\right]=0 .
\end{equation}
After rearranging the terms one obtains
\begin{equation}
\left(1-\frac{\xi b_0^2}{2}\right)\omega^2
-
\left(1+\frac{\xi b_0^2}{2}\right)\mathbf{k}^2
=0 .
\end{equation}
This relation can be written in the form
\begin{equation}
\label{omega}
\omega^2=v_t^{\,2}\,|\mathbf{k}|^2 ,
\end{equation}
where the effective propagation speed is
\begin{equation}
\label{v2}
v_t^{\,2}=
\frac{1+\frac{\xi b_0^2}{2}}
{1-\frac{\xi b_0^2}{2}} .
\end{equation}

In momentum space the Green function associated with this operator is therefore
\begin{equation}
\label{gomegap}
\widetilde{G}^{{(\text{t})}}(\omega,\mathbf{p})=
\frac{1}
{ { v_t^{\,2}\,|\mathbf{k}|^2 - \omega^2}},
\end{equation}
{where the label $(\text{t})$ denotes the timelike configuration.} Causal propagation is enforced by adopting the retarded prescription
\begin{equation}
\label{ieprescription}
\widetilde{G}^{{(\text{t})}}_{\rm ret}(\omega,\mathbf{p})=
\frac{1}{ { v_t^{\,2}\,|\mathbf{k}|^2 - \omega^2} -i\epsilon\,\omega}.
\end{equation}
{ The next step is to determine $\widetilde{G}^{{(\text{t})}}_{\rm ret}(t,\mathbf{k})$. For this purpose, we consider
\begin{equation}
G^{{(\text{t})}}_{\rm ret}(t,\mathbf{k})
=
\int_{-\infty}^{+\infty}
\frac{\mathrm{d}\omega}{2\pi}\,
e^{-i\omega t}\,
\widetilde{G}^{{(\text{t})}}_{\rm ret}(\omega,\mathbf{k}),
\end{equation}
and, using Eq.~(\ref{gomegap}), we obtain
\begin{equation}
G^{{(\text{t})}}_{\rm ret}(t,\mathbf{k})
=
\int_{-\infty}^{+\infty}
\frac{\mathrm{d}\omega}{2\pi}\,
\frac{e^{-i\omega t}}
{v_t^{\,2}\,|\mathbf{k}|^2 - \omega^2 -i\epsilon\,\omega}.
\end{equation}
For the sake of proceeding with the calculations, let us define $
\Omega_{\mathbf{k}}\equiv v_t|\mathbf{k}|$, so that
\begin{equation}
G^{{(\text{t})}}_{\rm ret}(t,\mathbf{k})
=
\int_{-\infty}^{+\infty}
\frac{\mathrm{d}\omega}{2\pi}\,
\frac{e^{-i\omega t}}
{\Omega_{\mathbf{k}}^2-\omega^2-i\epsilon\,\omega}.
\end{equation}
In the above case, the poles are determined by $\Omega_{\mathbf{k}}^2-\omega^2-i\epsilon\,\omega=0$.

Both poles lie below the real axis, as required by the retarded prescription. Therefore, closing the contour in the lower half plane for $t>0$, one obtains
\begin{equation}
G^{{(\text{t})}}_{\rm ret}(t,\mathbf{k})
=
\Theta(t)\,
\frac{\sin\left(\Omega_{\mathbf{k}}t\right)}
{\Omega_{\mathbf{k}}} = \Theta(t)\,
\frac{\sin\left(v_t|\mathbf{k}|t\right)}
{v_t|\mathbf{k}|},
\end{equation}
where $\Theta(t)$ denotes the Heaviside step function, defined as
\begin{equation}
\Theta(t)=
\begin{cases}
1, & t>0,\\
0, & t<0,
\end{cases}
\end{equation}
and its value at $t=0$ is conventional and does not affect the distributional Green function. In the present context, this factor enforces the retarded character of the solution, ensuring that the response vanishes for times prior to the action of the source.

Finally, in order to obtain the retartded Green's function into the coordinate space, through the expression below
\begin{equation}
\label{sseeeess}
G^{(\text{t})}_{\rm ret}(t,r)
=
\int\frac{\mathrm{d}^3k}{(2\pi)^3}
e^{i\mathbf{k}\cdot\mathbf{r}} G^{{(\text{t})}}_{\rm ret}(t,\mathbf{k}) =
\int\frac{\mathrm{d}^3k}{(2\pi)^3}
e^{i\mathbf{k}\cdot\mathbf{r}} \Theta(t)\,
\frac{\sin\left(v_t|\mathbf{k}|t\right)}
{v_t|\mathbf{k}|}.
\end{equation}    
To evaluate this integral, we use spherical coordinates in momentum space,
\begin{equation}
\mathrm{d}^3k=k^2\,\mathrm{d}k\,\mathrm{d}\Omega,
\qquad
\mathbf{k}\cdot\mathbf{r}=k \,r\cos\theta ,
\end{equation}
where, for notational simplicity, $r=|\mathbf r|$ and $k=|\mathbf k|$. The angular integration is then given by
\begin{equation}
\int \mathrm{d}\Omega\, e^{i\mathbf{k}\cdot\mathbf{r}}
=
2\pi\int_0^\pi \sin\theta\, e^{ikr\cos\theta}\,\mathrm{d}\theta
=
4\pi\,\frac{\sin(kr)}{kr}.
\end{equation}
Therefore,
\begin{align}
G^{{(\text{t})}}_{\rm ret}(t,r)
&=
\Theta(t)
\int_0^\infty
\frac{k^2\,\mathrm{d}k}{(2\pi)^3}
\left(
4\pi\,\frac{\sin(kr)}{kr}
\right)
\frac{\sin(v_tkt)}{v_t k} =
\frac{\Theta(t)}{2\pi^2 v_t r}
\int_0^\infty
\sin(kr)\sin(v_tkt)\,\mathrm{d}k .
\end{align}
Now, we use the distributional identity
\begin{equation}
\int_0^\infty \sin(kr)\sin(v_tkt)\,\mathrm{d}k
=
\frac{\pi}{2}
\bigg[
\delta(r-v_t t)-\delta(r+v_t t)
\bigg].
\end{equation}
Then, we have
\begin{equation}
G^{{(\text{t})}}_{\rm ret}(t,r)
=
\frac{\Theta(t)}{4\pi v_t r}
\bigg[
\delta(r-v_t t)-\delta(r+v_t t)
\bigg].
\end{equation}
Nevertheless, a physical retarded solution requires $r>0$ and $t>0$. In this manner, the term $\delta(r+v_t t)$ has no reasonability and must be discarded. Then,
\begin{equation}
G^{{(\text{t})}}_{\rm ret}(t,r)
=
\frac{\Theta(t)}{4\pi v_t r}\,
\delta(r-v_t t).
\end{equation}

Using the identity
\begin{equation}
\delta(r-v_t t)
=
\frac{1}{v_t}\,
\delta\left(t-\frac{r}{v_t}\right),
\end{equation}
we finally obtain
\begin{equation}
G^{{(\text{t})}}_{\rm ret}(t,r)
=
\frac{\Theta(t)}{4\pi v_t^2 r}\,
\delta\left(t-\frac{r}{v_t}\right).
\end{equation}
Equivalently, since the Dirac delta already enforces $t=r/v_t>0$, this can also be written as
\begin{equation}
G^{{(\text{t})}}_{\rm ret}(t,r)
=
\frac{1}{4\pi v_t^2 r}\,
\delta\left(t-\frac{r}{v_t}\right).
\label{gtr}
\end{equation}

Having obtained the retarded Green function $G^{{(\text{t})}}_{\rm ret}(t,r)$, displayed in Eq.~(\ref{gtr}), the metric perturbation, or strain, in the radiation zone follows from its retarded convolution with the spatial components of the stress--energy tensor. Therefore,
\begin{equation}
h_{ij}^{(\text{t}) \rm TT}(t,\mathbf{x})
=
16\pi G
\int_{-\infty}^{+\infty}\mathrm{d}t'
\int \mathrm{d}^{3}y\,
G^{{(\text{t})}}_{\rm ret}\left(t-t',|\mathbf{x}-\mathbf{y}|\right)
T_{ij}^{(\text{t}) \rm TT}(t',\mathbf{y}) .
\end{equation}

Using the explicit form of the Green function, we find
\begin{equation}
h_{ij}^{(\text{t}) \rm TT}(t,\mathbf{x})
=
16\pi G
\int_{-\infty}^{+\infty}\mathrm{d}t'
\int \mathrm{d}^{3}y\,
\frac{1}{4\pi v_t^2 |\mathbf{x}-\mathbf{y}|}
\delta\left(
t-t'
-
\frac{|\mathbf{x}-\mathbf{y}|}{v_t}
\right)
T_{ij}^{(\text{t}) \rm TT}(t',\mathbf{y}) .
\end{equation}

In the far--zone regime, where the observer is located at a distance $r=|\mathbf{x}|$ much larger than the size of the source, we are able to write
\begin{equation}
|\mathbf{x}-\mathbf{y}|\simeq r,
\qquad
\frac{1}{|\mathbf{x}-\mathbf{y}|}\simeq \frac{1}{r}.
\end{equation}
With these steps, we are able to write
\begin{equation}
h_{ij}^{(\text{t}) \rm TT}(t,\mathbf{x})
\simeq
\frac{16\pi G}{4\pi v_t^2 r}
\int_{-\infty}^{+\infty}\mathrm{d}t'
\delta\left(
t-t'
-
\frac{r}{v_t}
\right)
\int \mathrm{d}^{3}y\,
T_{ij}^{(\text{t}) \rm TT}(t',\mathbf{y}) .
\end{equation}

Defining the modified retarded time as
$t_r
\equiv
t-\frac{r}{v_t}$,
the integration over $t'$ gives
\begin{equation}
h_{ij}^{(\text{t}) \rm TT}(t,\mathbf{x})
\simeq
\frac{4G}{v_t^2 r}
\int \mathrm{d}^{3}y\,
T_{ij}^{(\text{t}) \rm TT}(t_r,\mathbf{y}) .
\end{equation}

For a localized and slowly moving source, the standard quadrupole identity gives \cite{d2022introducing}
\begin{equation}
\int \mathrm{d}^{3}y\,
T_{ij}^{(\text{t}) \rm TT}(t,\mathbf{y})
=
\frac{1}{2}
\frac{\mathrm{d}^{2}I_{ij}^{(\text{t}) \rm TT}}{\mathrm{d}t^{2}}(t).
\label{identitityquadripole}
\end{equation}
where the mass quadrupole moment is defined by
\begin{equation}
I_{ij}(t)
=
\int \mathrm{d}^{3}y\,
y_i y_j T^{00}(t,\mathbf{y}),
\end{equation}
so that the associated strain is
\begin{equation}
\label{straintime}
h_{ij}^{(\text{t}) \rm TT}(t,r)
=
\frac{2G}{v_t^2 r}
\frac{\mathrm{d}^{2}I_{ij}^{(\text{t}) \rm TT}}{\mathrm{d}t^{2}}
\left(t-\frac{r}{v_t}\right).
\end{equation} }
The waveform thus retains the usual quadrupole dependence on the source dynamics, while the retarded time and the overall amplitude {are modified through their explicit dependence on $v_t$, which contains the Lorentz--violating contribution, as shown in Eq.~(\ref{v2})}.


\subsection{The spacelike case}

Following the strategy adopted in Ref.~\cite{AraujoFilho:2026zyt} and in the bumblebee wave analysis of Ref.~\cite{Amarilo:2023wpn}, we now address the \emph{spacelike} configuration of the background vector in the \textit{metric--affine} bumblebee scenario, starting from the modified dispersion relation in Eq. (\ref{disp_grav_eikonal}). We choose a purely spacelike vacuum expectation value,
$b^\mu=(0,\mathbf{b}),\, b^2=-|\mathbf{b}|^2,
\,\, b\!\cdot\!k=-\,\,\mathbf{b}\!\cdot\!\mathbf{k}=-|\mathbf{b}|\,|\mathbf{k}|\cos\Psi$,
where $\Psi$ is the angle between $\mathbf{k}$ and $\mathbf{b}$. We parametrize this angle through the spherical directions of $\mathbf{k}$ and $\mathbf{b}$,
$\cos\Psi=\cos\theta\,\cos\chi+\sin\theta\,\sin\chi\cos(\phi_b-\phi)$,
with $(\theta,\phi)$ and $(\chi,\phi_b)$ the angular coordinates of $\mathbf{k}$ and $\mathbf{b}$, respectively.

Substituting $b^2=-|\mathbf{b}|^2$ and $(b\!\cdot\!k)^2=|\mathbf{b}|^2|\mathbf{k}|^2\cos^2\Psi$ into Eq.~\eqref{disp_grav_eikonal} yields
\begin{align}
0
&=(\omega^2-|\mathbf{k}|^2)-\xi\left[|\mathbf{b}|^2|\mathbf{k}|^2\cos^2\Psi-\frac12(-|\mathbf{b}|^2)(\omega^2-|\mathbf{k}|^2)\right]
\nonumber\\
&=\left(1-\frac{\xi|\mathbf{b}|^2}{2}\right)\omega^2
-\left(1-\frac{\xi|\mathbf{b}|^2}{2}+\xi|\mathbf{b}|^2\cos^2\Psi\right)|\mathbf{k}|^2 .
\label{MA_MDR_spacelike_explicit}
\end{align}
{Then}, up to $\mathcal O(\xi)$ the dispersion law can be written as
\begin{equation}
\omega^2=v_s^{\,2}(\Psi)\,|\mathbf{k}|^2,
\qquad
v_s^{\,2}(\Psi)=
\frac{1-\frac{\xi|\mathbf{b}|^2}{2}+\xi|\mathbf{b}|^2\cos^2\Psi}{1-\frac{\xi|\mathbf{b}|^2}{2}}.
\label{vs_spacelike}
\end{equation}

In momentum space, the Green function associated with the modified wave operator reads
\begin{equation}
\widetilde G^{\text{(s)}}(\omega,\mathbf{k})=
\frac{1}{ { v_s^{\,2}(\Psi)\,|\mathbf{k}|^2 - \omega^2}},
\qquad
\cos\Psi=\frac{\mathbf{b}\!\cdot\!\mathbf{k}}{|\mathbf{b}|\,|\mathbf{k}|},
\label{Gtilde_spacelike_exact}
\end{equation}
and causal propagation is fixed by the retarded prescription {similar to we used for the timelike case in Eq. (\ref{ieprescription})}. {In this manner, the retarded Green function is obtained through the inverse Fourier transform
\begin{equation}
G^{{(\text{s})}}_{\rm ret}(t,\mathbf{k})
=
\int_{-\infty}^{+\infty}
\frac{\mathrm{d}\omega}{2\pi}\,
\frac{e^{-i\omega t}}
{
v_s^{\,2}(\Psi)|\mathbf{k}|^2
-\omega^2
-i\epsilon\omega
}.
\end{equation}
The term $-i\epsilon\omega$ implements the retarded prescription, shifting the poles below the real axis in the appropriate way and ensuring causal support for $t>0$. The poles are located at
\begin{equation}
\omega_{\pm}
\simeq
\pm v_s(\Psi)|\mathbf{k}|
-i\epsilon .
\end{equation}
Then, closing the contour in the lower half--plane for $t>0$, one obtains
\begin{equation}
G^{{(\text{s})}}_{\rm ret}(t,\mathbf{k})
=
\Theta(t)\,
\frac{
\sin\left[v_s(\Psi)|\mathbf{k}|t\right]
}{
v_s(\Psi)|\mathbf{k}|
}.
\label{sdeerrr}
\end{equation} 
As we have seen up to now, the methodology for calculating the Green's function $G^{{(\text{t})}}_{\rm ret}(t,|\mathbf{k}|)$ and $G^{{(\text{s})}}_{\rm ret}(t,|\mathbf{k}|)$  are so similar. However, for evaluating $G^{{(\text{s})}}_{\rm ret}(t,r)$, there will appear some delicate steps for the spacelike configuration due to the anisotropy bring about by the dependence of the angle $\Psi$, as we shall be seeing below.

}

{ Analogous to what we have done for the timelike case in Eq. (\ref{sseeeess}), by using Eq. (\ref{sdeerrr}), we can write
\begin{align}
G^{{(\text{s})}}_{\rm ret}(t,\mathbf r)
&=
\Theta(t)
\int
\frac{\mathrm d^3 k}{(2\pi)^3}
e^{i\mathbf{k}\cdot\mathbf r}
\frac{
\sin\left[
t\sqrt{|\mathbf{k}|^2+\gamma(\hat{\mathbf b}\cdot\mathbf{k})^2}
\right]
}{
\sqrt{|\mathbf{k}|^2+\gamma(\hat{\mathbf b}\cdot\mathbf{k})^2}
},
\end{align}
where
\begin{equation}
\gamma
\equiv
\frac{\xi|\mathbf b|^2}
{1-\frac{\xi|\mathbf b|^2}{2}} .
\end{equation}
Choosing the axis parallel to the spacelike background vector $\mathbf b$, one has
\begin{equation}
\hat{\mathbf b}\cdot\mathbf{k}=k_{\parallel},
\qquad
|\mathbf{k}|^2=k_{\perp}^{2}+k_{\parallel}^{2},
\end{equation}
and therefore
\begin{equation}
|\mathbf{k}|^2+\gamma(\hat{\mathbf b}\cdot\mathbf{k})^2
=
k_{\perp}^{2}+\beta k_{\parallel}^{2},
\qquad
\beta\equiv 1+\gamma
=
\frac{1+\frac{\xi|\mathbf b|^2}{2}}
{1-\frac{\xi|\mathbf b|^2}{2}} .
\end{equation}
Introducing the rescaled momentum component
\begin{equation}
q_{\parallel}=\sqrt{\beta}\,k_{\parallel},
\qquad
q_{\perp}=k_{\perp},
\end{equation}
we obtain
\begin{equation}
\mathrm d^3 k
=
\frac{\mathrm d^3 q}{\sqrt{\beta}},
\qquad
\mathbf{k}\cdot\mathbf r
=
\mathbf q_{\perp}\cdot\mathbf r_{\perp}
+
q_{\parallel}\frac{r_{\parallel}}{\sqrt{\beta}} .
\end{equation}
In this case, notice that the inverse Fourier transform reduces to the standard retarded Green function evaluated at the deformed distance
\begin{equation}
R_s
=
\sqrt{
r_{\perp}^{2}
+
\frac{r_{\parallel}^{2}}{\beta}
}.
\end{equation}
Then,
\begin{equation}
G^{{(\text{s})}}_{\rm ret}(t, r)
=
\frac{\Theta(t)}
{4\pi\sqrt{\beta}\,R_s}
\,
\delta(t-R_s).
\end{equation}
If $\chi$ denotes the angle between $\mathbf r$ and $\mathbf b$, then
\begin{equation}
r_{\parallel}=r\cos\chi,
\qquad
r_{\perp}=r\sin\chi,
\end{equation}
so that
\begin{equation}
R_s
=
r
\sqrt{
\sin^2\chi+\frac{\cos^2\chi}{\beta}
}.
\end{equation}
Up to first order in $\xi$, one has
\begin{equation}
\beta
=
1+\xi|\mathbf b|^2
+
\mathcal O(\xi^2),
\end{equation}
and therefore
\begin{equation}
R_s
=
r
\left[
1-\frac{\xi|\mathbf b|^2}{2}\cos^2\chi
\right]
+
\mathcal O(\xi^2),
\end{equation}
while
\begin{equation}
\frac{1}{\sqrt{\beta}\,R_s}
=
\frac{1}{r}
\left[
1-\frac{\xi|\mathbf b|^2}{2}\sin^2\chi
\right]
+
\mathcal O(\xi^2).
\end{equation}
Moreover,
\begin{align}
\delta(t-R_s)
&=
\delta\left(
t-r+\frac{\xi|\mathbf b|^2}{2}r\cos^2\chi
\right)
+
\mathcal O(\xi^2)
\nonumber\\
&=
\delta(t-r)
+
\frac{\xi|\mathbf b|^2}{2}r\cos^2\chi\,
\delta'(t-r)
+
\mathcal O(\xi^2).
\end{align}
Consequently, in an elegant manner, we can obtain the retarded Green function in coordinate space is
\begin{equation}
G^{{(\text{s})}}_{\rm ret}(t,\mathbf r)
=
\frac{\Theta(t)}{4\pi r}
\left\{
\left[
1-\frac{\xi|\mathbf b|^2}{2}\sin^2\chi
\right]\delta(t-r)
+
\frac{\xi|\mathbf b|^2}{2}\,
r\cos^2\chi\,
\delta'(t-r)
\right\}
+
\mathcal O(\xi^2).
\end{equation}

Using the standard quadrupole identity presented in Eq.~\eqref{identitityquadripole}, we obtain
\begin{equation}
\int \mathrm{d}t'\,
\delta(t-t'-r)
\int \mathrm{d}^{3}y\,T_{ij}(t',\mathbf y)
=
\frac{1}{2}
\frac{\mathrm{d}^{2} I_{ij}}{\mathrm{d}t^{2}}(t^{\text{s}}_r),
\end{equation}
and
\begin{equation}
\int \mathrm{d}t'\,
\delta'(t-t'-r)
\int \mathrm{d}^{3}y\,T_{ij}(t',\mathbf y)
=
\frac{1}{2}
\frac{\mathrm{d}^{3} I_{ij}}{\mathrm{d}t^{3}}(t^{\text{s}}_r),
\end{equation}
where $t^{\text{s}}_r=t-r$. Therefore, after the transverse--traceless projection, the spatial metric perturbation becomes
\begin{equation}
h_{ij}^{(\mathrm{s})\,\mathrm{TT}}(t,\mathbf r)
=
\frac{2G}{r}
\left\{
\left[
1-\frac{\xi|\mathbf b|^2}{2}\sin^2\chi
\right]
\frac{\mathrm{d}^{2} I_{ij}^{\mathrm{TT}}}{\mathrm{d}t^{2}}(t^{\text{s}}_r)
+
\frac{\xi|\mathbf b|^2}{2}\,
r\cos^2\chi\,
\frac{\mathrm{d}^{3} I_{ij}^{\mathrm{TT}}}{\mathrm{d}t^{3}}(t^{\text{s}}_r)
\right\}
+
\mathcal O(\xi^2).
\label{rzrrr}
\end{equation}

}

Therefore, in the spacelike sector the Lorentz--violating background does not reduce to a retarded time shift. Instead, it introduces a directional dependence through $\chi$ together with higher--derivative radiative corrections.


\section{Gravitational radiation from a binary black hole}

\begin{figure}
    \centering
    \includegraphics[scale=0.52]{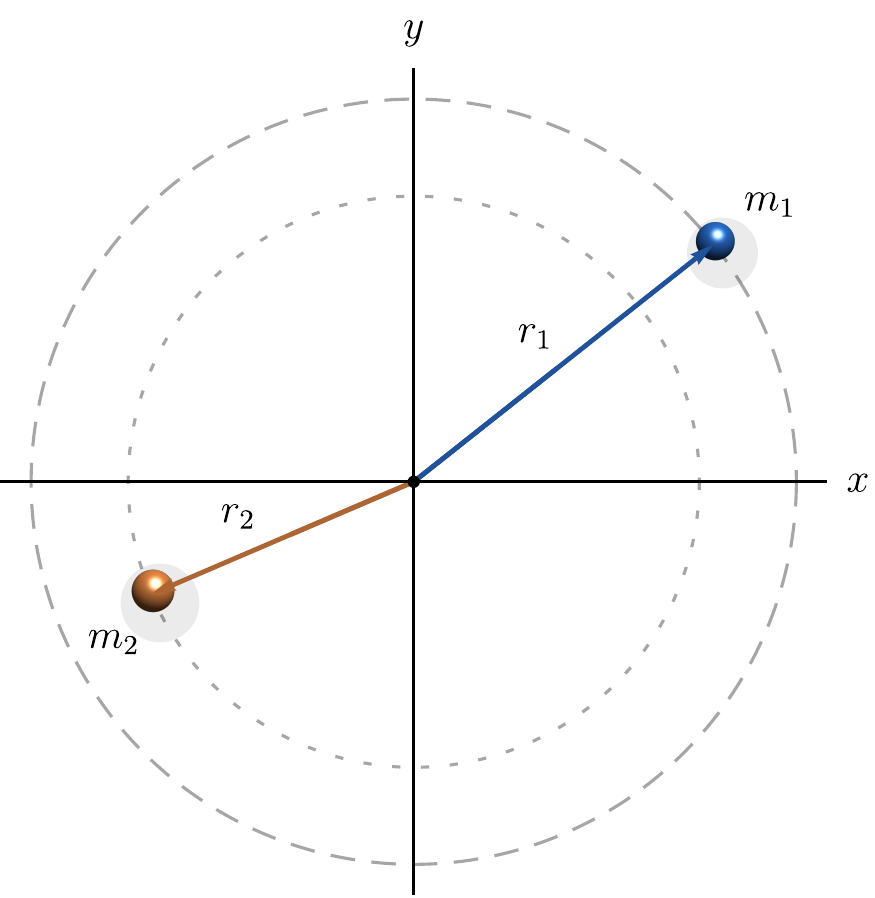}
    \caption{Binary system viewed in the center of mass frame, where two compact objects with masses $m_1$ and $m_2$ move in circular orbits on the $xy$ plane with radii $r_1$ and $r_2$.}
    \label{binary}
\end{figure}

To illustrate the consequences of the Lorentz--violating propagation law, we consider gravitational radiation generated by a binary black hole system, as shown in Fig.~\ref{binary}. Two compact objects with masses $m_1$ and $m_2$ orbit in the $xy$ plane. The motion is described in the center of mass frame, where the positions of the bodies are measured relative to this point, with orbital radii $r_1$ and $r_2$.

The system is approximated as two point masses restricted to the orbital plane. In this representation, the matter distribution is encoded in the energy density component of the stress--energy tensor,
\ie
T_{00}=\delta(z)\big[m_{1}\delta(x-x_{1})\delta(y-y_{1})+m_{2}\delta(x-x_{2})\delta(y-y_{2})\big],
\fe
which concentrates the source at the instantaneous positions of the two bodies.

For circular motion in the center of mass frame, the trajectories of the two bodies can be written as
\ie
x_{1}(t)=\frac{m_{2}l_{0}}{M}\cos(\omega t), \quad 
y_{1}(t)=\frac{m_{2}l_{0}}{M}\sin(\omega t),
\quad x_{2}(t)=-\frac{m_{1}l_{0}}{M}\cos(\omega t),
\quad y_{2}(t)=-\frac{m_{1}l_{0}}{M}\sin(\omega t),
\fe
where $M=m_1+m_2$ is the total mass. The constant separation between the components is $l_0=r_1+r_2$, and $\omega$ denotes the orbital frequency.

For this binary configuration, the mass quadrupole tensor simplifies considerably, leaving only three independent components different from zero
\ie
\begin{split}
\label{compI}
I_{xx}(t)
&=
\frac{\mu}{2}l_{0}^{2}\big[1+\cos(2\omega t)\big], \,\, I_{yy}(t)
=
\frac{\mu}{2}l_{0}^{2}\big[1-\cos(2\omega t)\big], \, \, 
I_{xy}(t)
=
I_{yx}(t)
=
\frac{\mu}{2}l_{0}^{2}\sin(2\omega t).
\end{split}
\fe
Here, $\mu=\frac{m_1 m_2}{m_1+m_2}$ represents the reduced mass of the binary system.

\subsection{Timelike configuration}

Using the circular--binary quadrupole components given in Eq.~\eqref{compI}, together with the radiation--zone result in Eq.~\eqref{straintime}, the waveform in the timelike background {can be obtained}. The nonvanishing components are therefore
\begin{align}
h_{xx}^{(\text{t})}(t,r)
&=
-\frac{4G\,\mu\,l_{0}^{2}\,\omega^{2}}
{{v_{t}^{2}}\,r}\,
\cos\!\Big(2\omega\,t_{r}\Big), \label{hxxt}\\
h_{yy}^{(\text{t})}(t,r)
&=
+\frac{4G\,\mu\,l_{0}^{2}\,\omega^{2}}
{{v_{t}^{2}}\,r}\,
\cos\!\Big(2\omega\,t_{r}\Big), \\
h_{xy}^{(\text{t})}(t,r)=h_{yx}^{(\text{t})}(t,r)
&=
-\frac{4G\,\mu\,l_{0}^{2}\,\omega^{2}}
{{v_{t}^{2}}\,r}\,
\sin\!\Big(2\omega\,t_{r}\Big).
\end{align}
For a face--on observer, the transverse--traceless polarizations are identified as
$h_{+}^{(\text{t})}\equiv h_{xx}^{(\text{t})}=-h_{yy}^{(\text{t})}$
and
$h_{\times}^{(\text{t})}\equiv h_{xy}^{(\text{t})}$.
Therefore, in the timelike configuration, Lorentz violation modifies the waveform through two effects: the retarded time is shifted by the effective propagation speed $v_t$, and the overall amplitude is rescaled by the factor ${v_{t}^{-2}}$. The polarization pattern, however, remains the same as in the standard quadrupole result.

\begin{figure}
    \centering
    \includegraphics[scale=0.55]{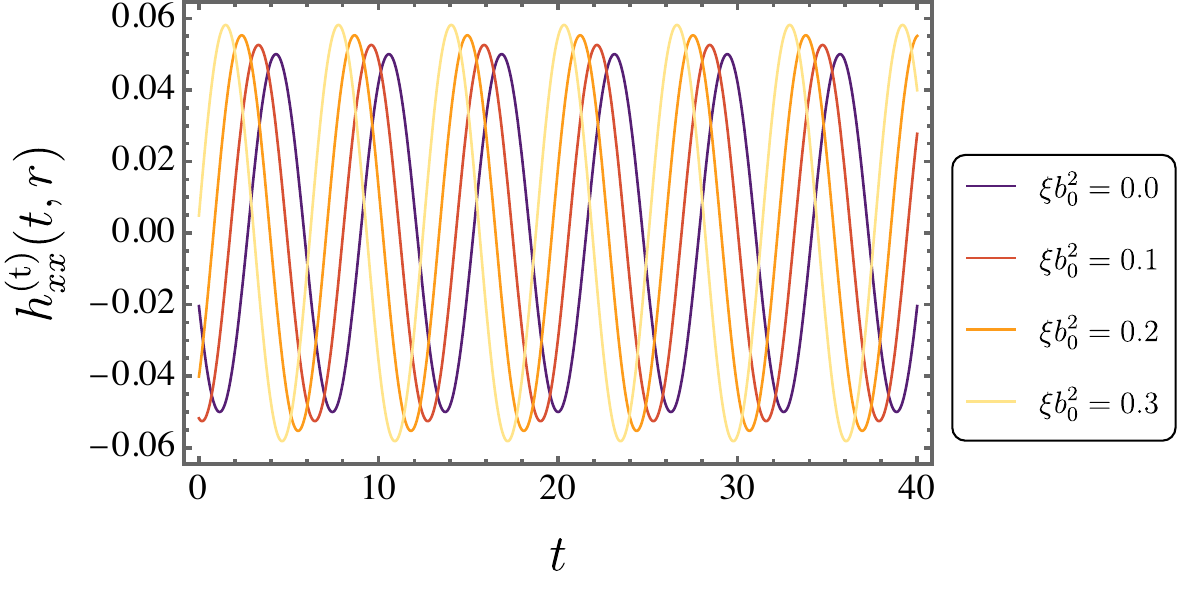}
    \caption{Time evolution of the gravitational wave component $h_{xx}^{(\text{t})}(t,r)$ for the timelike configuration, plotted for $\xi b_0^2={0.0,0.1,0.2,0.3}$ with $\omega=0.5$, $r=20$, $\mu=1$, $l_0=1$, and $G=1$.}
    \label{hxxtimilike}
\end{figure}

In Fig.~\ref{hxxtimilike}, we display the time evolution of the gravitational wave component $h_{xx}^{(t)}(t,r)$ for the timelike configuration of the Lorentz--violating background. The curves correspond to the values $\xi b_0^2=\{0.0,\,0.1,\,0.2,\,0.3\}$, with $\omega=0.5$, $r=20$, $\mu=1$, $l_0=1$, and $G=1$. As $\xi b_0^2$ increases, the effective propagation speed $v_t$ changes, which shifts the phase of the oscillatory signal through the modified retarded time $t_r$. At the same time, the amplitude is slightly suppressed by the factor ${v_{t}^{-2}}$, {as mentioned before. In addition, increasing the Lorentz--violating coefficient $\xi b_{0}$ enhances the strain amplitude and shifts the waveform phase toward earlier times.

} 

\subsection{Spacelike configuration}

In the spacelike background, the radiation zone field \eqref{rzrrr} exhibits two distinct Lorentz--violating structures: first, an anisotropic rescaling of the standard quadrupole piece and, second, an additional contribution proportional to $\dddot I_{ij}$.

Since the signal propagates from the source to the observer along $\hat{\mathbf r}$, the relevant wave--vector direction satisfies $\hat{\mathbf k}=\hat{\mathbf r}$. Therefore, the angle entering the spacelike dispersion relation reduces to the angle between the background vector $\mathbf b$ and the line of sight,
$\Psi=\chi,
\,\,
\mathbf b\!\cdot \mathbf r = |\mathbf b|\,r\cos\chi$.

Specializing to the circular--binary quadrupole components \eqref{compI}, we need
\begin{align}
\ddot I_{xx}(t)=-2\mu l_0^2\omega^2\cos(2\omega t),
&\qquad
\dddot I_{xx}(t)=+4\mu l_0^2\omega^3\sin(2\omega t),\nonumber\\
\ddot I_{yy}(t)=+2\mu l_0^2\omega^2\cos(2\omega t),
&\qquad
\dddot I_{yy}(t)=-4\mu l_0^2\omega^3\sin(2\omega t),\nonumber\\
\ddot I_{xy}(t)=-2\mu l_0^2\omega^2\sin(2\omega t),
&\qquad
\dddot I_{xy}(t)=-4\mu l_0^2\omega^3\cos(2\omega t).
\end{align}

In this manner, these equations above together with Eq. (\ref{rzrrr}), we have
\begin{align}
h_{xx}^{(\text{s})}(t,r)
=&
-\frac{4G\,\mu\,l_{0}^{2}\,\omega^{2}}{r}
\left(1-\frac{\xi|\mathbf b|^2}{2}\sin^{2}\chi\right)
\cos\!\big(2\omega t^{\text{s}}_r\big) \\
& 
+4G\,\mu\,l_{0}^{2}\,\omega^{3}\,\xi|\mathbf b|^{2}\cos^{2}\chi\,
\sin\!\big(2\omega t^{\text{s}}_r\big),
\\[1mm]
h_{yy}^{(\text{s})}(t,r)
=&
+\frac{4G\,\mu\,l_{0}^{2}\,\omega^{2}}{r}
\left(1-\frac{\xi|\mathbf b|^2}{2}\sin^{2}\chi\right)
\cos\!\big(2\omega t^{\text{s}}_r\big) \\
& 
-4G\,\mu\,l_{0}^{2}\,\omega^{3}\,\xi|\mathbf b|^{2}\cos^{2}\chi\,
\sin\!\big(2\omega t^{\text{s}}_r\big),
\\[1mm]
h_{xy}^{(\text{s})}(t,r)=h_{yx}^{(\text{s})}(t,r)
=&
-\frac{4G\,\mu\,l_{0}^{2}\,\omega^{2}}{r}
\left(1-\frac{\xi|\mathbf b|^2}{2}\sin^{2}\chi\right)
\sin\!\big(2\omega t^{\text{s}}_r\big) \\
&
-4G\,\mu\,l_{0}^{2}\,\omega^{3}\,\xi|\mathbf b|^{2}\cos^{2}\chi\,
\cos\!\big(2\omega t^{\text{s}}_r\big).
\end{align}

As it happened in the timelike configuration analysis, for a face on observer, the TT polarizations satisfy
$h_{+}^{(\text{s})}\equiv h_{xx}^{(\text{s})}=-h_{yy}^{(\text{s})}$ and $h_{\times}^{(\text{s})}\equiv h_{xy}^{(\text{s})}$.

Therefore, in the spacelike \textit{metric--affine} bumblebee sector, Lorentz violation produces a direction--dependent modulation of the overall quadrupole amplitude through the combination
$1-\frac{\xi|\mathbf b|^2}{2}\sin^{2}\chi$, and a polarization--mixing correction proportional to $\omega^3\xi|\mathbf b|^2\cos^2\chi$, arising from the $\dddot I_{ij}$ term.

\begin{figure}
    \centering
    \includegraphics[scale=0.55]{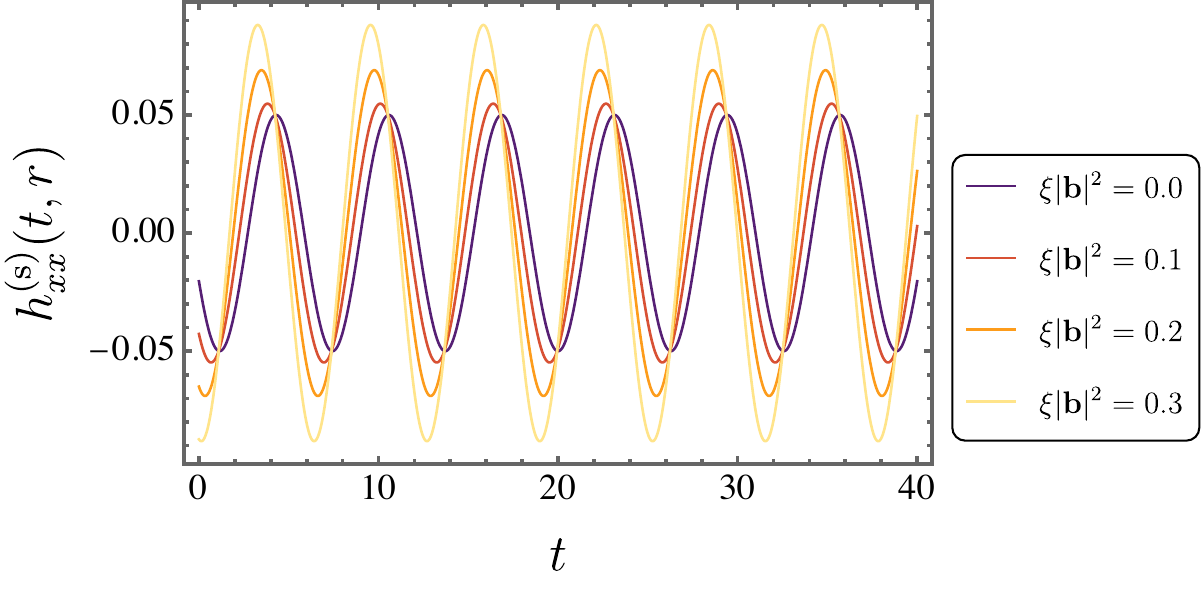}
    \caption{Time evolution of the gravitational wave component $h_{xx}^{(\text{s})}(t,r)$ for the spacelike configuration of the Lorentz--violating background. The curves correspond to $\xi|\mathbf b|^2={0.0,0.1,0.2,0.3}$, with fixed parameters $\omega=0.5$, $r=20$, $\mu=1$, $l_0=1$, $G=1$, and $\chi=\pi/4$.}
    \label{hxxspacelike}
\end{figure}

In Fig.~\ref{hxxspacelike}, we display the time evolution of the gravitational wave component $h_{xx}^{(\text{s})}(t,r)$ for the spacelike configuration of the background vector. The curves correspond to different values of the parameter $\xi|\mathbf b|^2={0.0,0.1,0.2,0.3}$, while the remaining parameters are fixed to $\omega=0.5$, $r=20$, $\mu=1$, $l_0=1$, $G=1$, and $\chi=\pi/4$. As the Lorentz--violating parameter increases, the waveform progressively departs from the general--relativistic profile due to the anisotropic corrections proportional to $\sin^{2}\chi$ and $\cos^2\chi$, which modify both the amplitude and the phase of the oscillations. {Furthermore, we see that the corresponding plot shows that increasing the Lorentz--violating parameter $\xi |\mathbf{b}|^{2}$ enhances the waveform amplitude and shifts the profile to the left.  }

\begin{figure}
    \centering
    \includegraphics[scale=0.43]{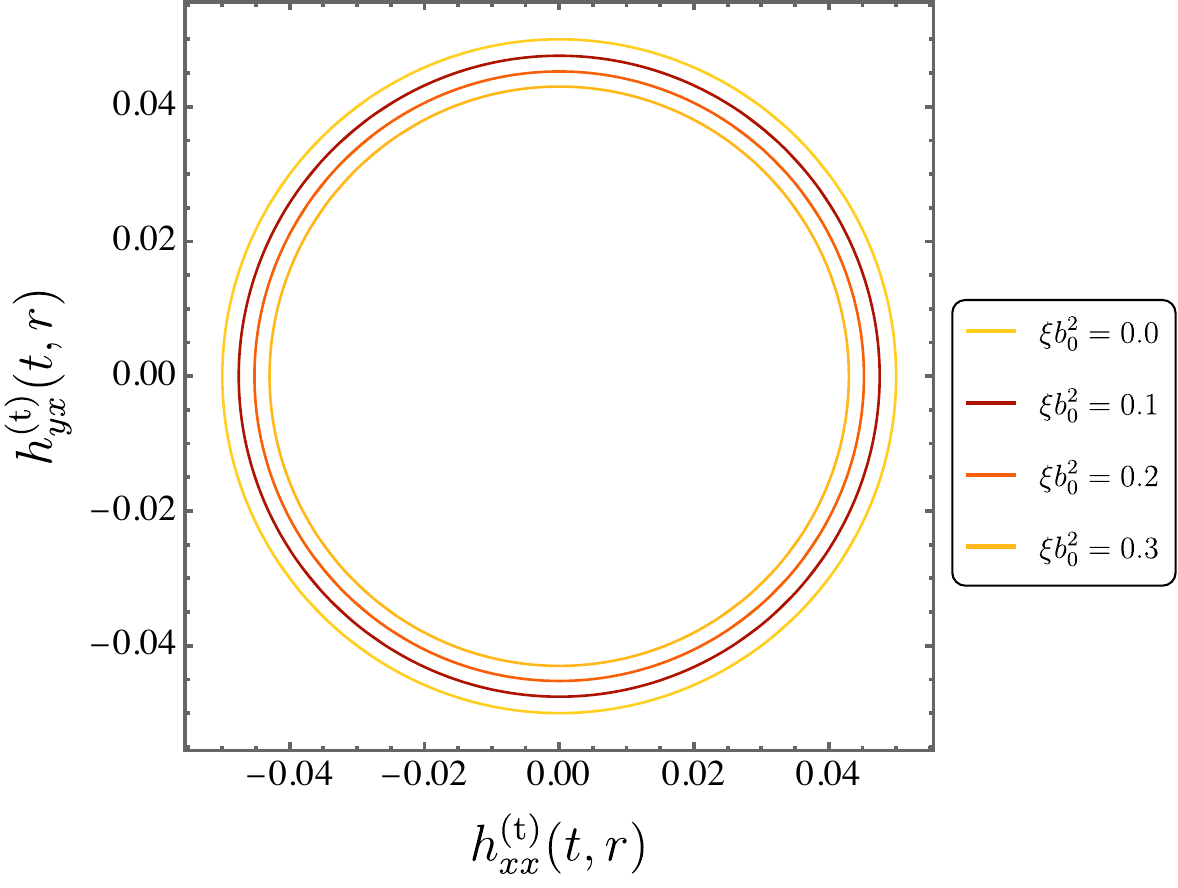}
    \includegraphics[scale=0.43]{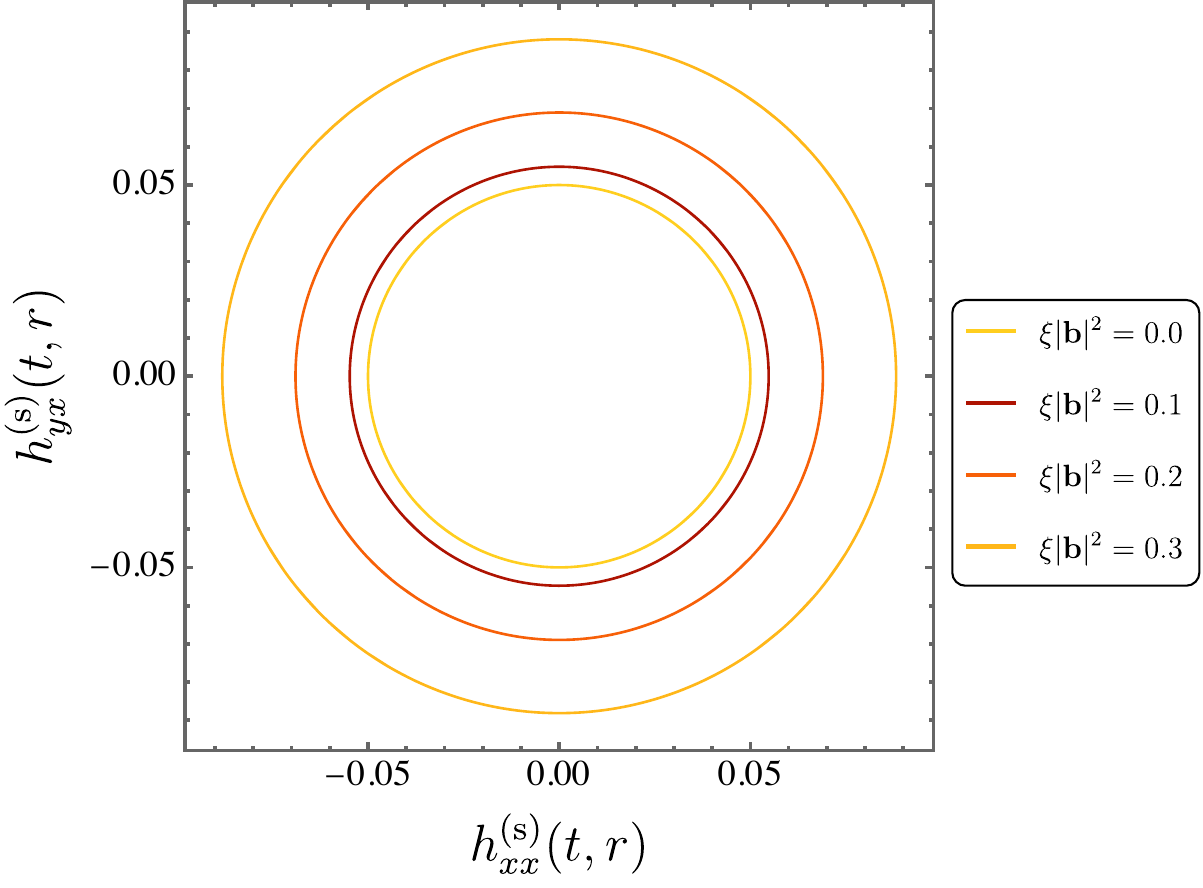}
    \caption{Parametric trajectories of the gravitational wave strains in the $(h_{xx},h_{yx})$ plane for a circular binary system, shown for timelike (on the left panel) and spacelike (on the right panel) configurations of the background vector $b^{\mu}$ and different values of $\xi b^{2}$. }
    \label{paramtericcc}
\end{figure}

In Fig.~\ref{paramtericcc}, we present the parametric trajectories of the strains $h_{yx}$ versus $h_{xx}$ for both timelike and spacelike configurations of the background vector $b^{\mu}$.

\begin{figure}
    \centering
    \includegraphics[scale=0.45]{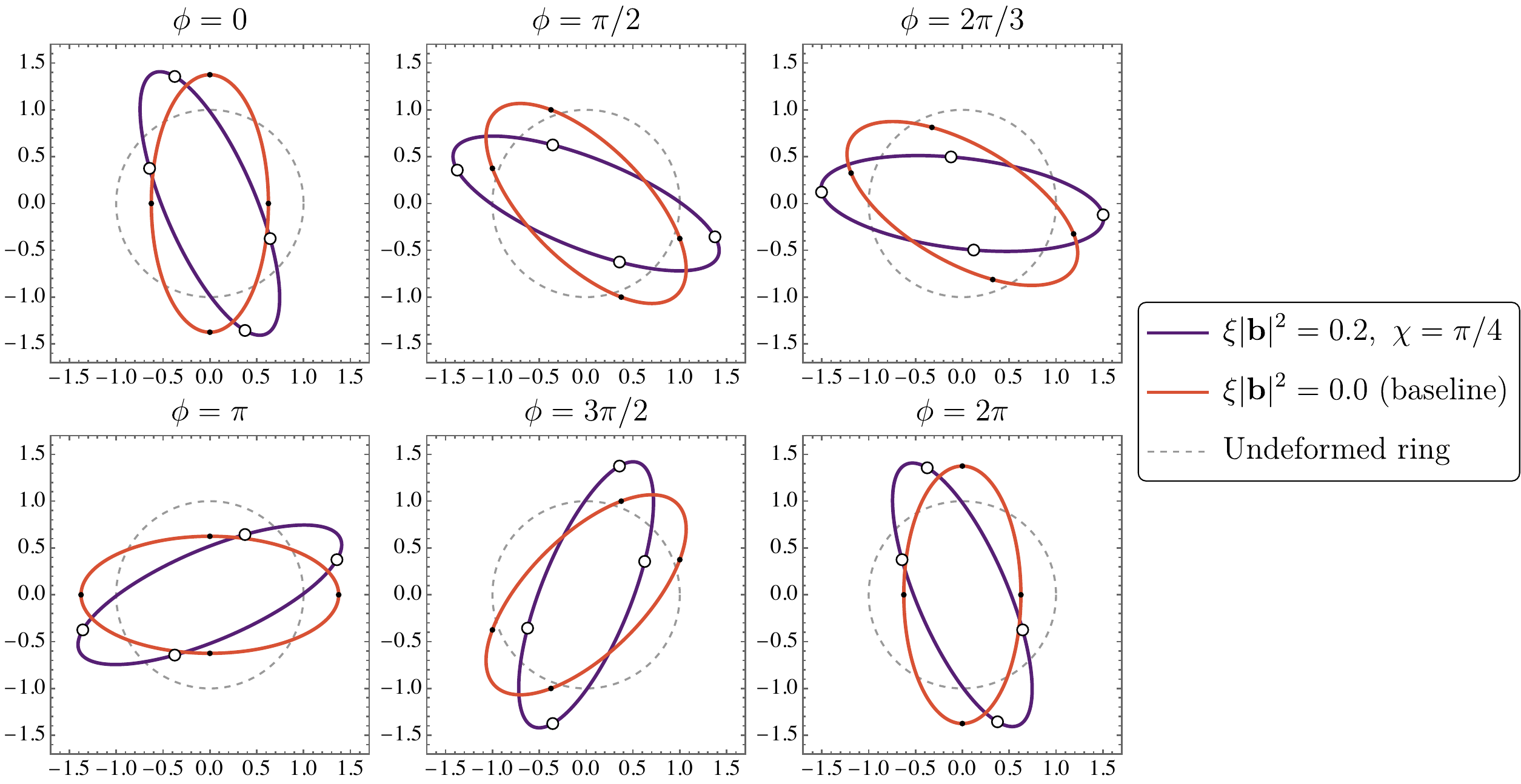}
    \caption{Ring deformation induced by a spacelike \textit{metric--affine} bumblebee gravitational wave. Six snapshots of the initially circular ring are shown at phases $\phi={0,\pi/2,2\pi/3,\pi,3\pi/2,2\pi}$, comparing the Lorentz--violating case ($\xi|\mathbf b|^2=0.2$, $\chi=\pi/4$) with the GR baseline ($\xi|\mathbf b|^2=0$); the dashed circle denotes the undeformed ring.}
    \label{polarizations}
\end{figure}

Fig. \ref{polarizations} shows the deformation of an initially circular ring of test particles produced by the passage of the gravitational wave in the spacelike configuration of the \textit{metric--affine} bumblebee model. The six panels correspond to successive phases $\phi={0,\pi/2,2\pi/3,\pi,3\pi/2,2\pi}$, illustrating the time evolution of the transverse displacement. The dashed circle denotes the undeformed ring, while the black points represent the GR prediction obtained for $\xi|\mathbf b|^2=0$. The colored curve corresponds to the Lorentz--violating case with $\xi|\mathbf b|^2=0.2$ and $\chi=\pi/4$. In GR the deformation is purely quadrupolar, producing the familiar alternating stretching and compression of the ring. In the spacelike Lorentz--violating scenario the deformation pattern is modified. The quadrupole amplitude is anisotropically rescaled and an additional phase--shifted contribution appears in the strain. Consequently, the ring deformation slightly departs from the GR pattern, producing small rotations and distortions of the elliptical shapes as the wave evolves.

\section{Constraints from gravitational wave observations}\label{sec:bounds}

In this section we estimate conservative constraints on the Lorentz--violating combination $\xi b^2$ in the \textit{metric--affine} bumblebee scenario using published gravitational wave observations. Our strategy is twofold: first, translate multimessenger limits into bounds on an effective GW speed whenever the propagation law can be written as $\omega^2=v^2 k^2$, and, second, impose \emph{strain--consistency} requirements directly on the radiation--zone waveform whenever new radiative structures appear. No detector data are reanalyzed here.

\subsection{Propagation bounds from GW170817/GRB\,170817A}\label{subsec:prop_bounds}

\subsubsection{Timelike background}

For $b^\mu=(b_0,0,0,0)$, the tensor modes propagate with $\omega^2=v_t^{\,2}|\mathbf{k}|^2$ and,
for $|\xi b_0^2|\ll 1$, one {obtains}
\begin{equation}
v_t \simeq 1+\frac{\xi}{2}\,b_0^{\,2}
\qquad\Longrightarrow\qquad
\frac{v_t-c}{c} \simeq \frac{\xi}{2}\,b_0^{\,2},
\label{dv_timelike}
\end{equation}
{where $c=1$ is used in the dispersion relation and restored only in the observational ratio.}
The joint detection of GW170817 and GRB\,170817A constrains the {relative} difference between the speed of
gravity and the speed of light to lie in the interval
$-3\times 10^{-15}\lesssim (v_g-c)/c \lesssim 7\times 10^{-16}$,
as inferred from the observed $\sim 1.7\,{\rm s}$ delay between the GW merger time and the onset
of gamma rays~\cite{Abbott:2017GWGRB}.
{Using the most conservative edge of this interval,} a conservative symmetric bound is therefore
\begin{equation}
\left|\frac{v_g-c}{c}\right|\lesssim 3\times 10^{-15}
\qquad\Rightarrow\qquad
|\xi|\,b_0^{\,2}\ \lesssim\ 6\times 10^{-15}.
\label{bound_timelike}
\end{equation}
{Equivalently, keeping the asymmetric interval, one would obtain the sign-sensitive estimate}
\begin{equation}
{
-6\times 10^{-15}\lesssim \xi b_0^{\,2}\lesssim 1.4\times 10^{-15}.
}
\label{bound_timelike_asymmetric}
\end{equation}

\subsubsection{Spacelike background parallel to propagation}

For $b^\mu=(0,0,0,b_3)$ with propagation along the $z$ axis, the dispersion relation again takes
the form $\omega^2=v_s^{\,2}|\mathbf{k}|^2$ and yields the same linearized mapping
{
\[
\frac{v_s-c}{c}\simeq \frac{\xi b_3^2}{2}
\]
}
in the perturbative regime. Hence, {using the conservative symmetric form of} the multimessenger bound {one obtains}
\begin{equation}
|\xi|\,b_3^{\,2}\ \lesssim\ 6\times 10^{-15}
\qquad
\text{(spacelike background aligned with propagation).}
\label{bound_spacelike_parallel}
\end{equation}
{If the asymmetric GW170817/GRB\,170817A interval is retained instead, the corresponding sign-sensitive estimate is}
\begin{equation}
{
-6\times 10^{-15}\lesssim \xi b_3^{\,2}\lesssim 1.4\times 10^{-15}.
}
\label{bound_spacelike_parallel_asymmetric}
\end{equation}
{For a generic spacelike orientation, the same argument constrains the projected combination
$|\xi|\,|\mathbf b|^2\cos^2\Psi$ rather than $|\xi|b_3^2$.}

{

\subsection{Strain--based bounds from waveform consistency}\label{subsec:strain_bounds}

Propagation bounds do not exhaust the available information because the radiation--zone waveform is also modified at the level of the strain. In the timelike configuration, the waveform receives an overall amplitude renormalization and a shifted retarded time. In the spacelike configuration, the waveform receives a direction--dependent amplitude modulation and an additional contribution proportional to $\dddot I_{ij}$. Since LVK signals are accurately described by GR templates, a conservative requirement is that Lorentz--violating corrections remain below the characteristic fractional waveform uncertainty across the observed frequency band. The estimates below should be interpreted as consistency bounds obtained at the waveform level, without reanalyzing detector data.


\subsubsection{Choice of tolerances from observational systematics}

Two observationally motivated tolerance scales may be identified. A first one is a \emph{detection-limited} tolerance of order $1/\rho$, where $\rho$ denotes the network matched-filter signal-to-noise ratio (SNR). For GW170817 the reported network value is $\rho\simeq 32.4$ \cite{Abbott:2017GW170817PRL}, implying $1/\rho\simeq 3\times 10^{-2}$. A second relevant scale is associated with \emph{calibration uncertainties} of the detectors, which typically correspond to amplitude variations at the level of a few percent and phase uncertainties of order $\mathcal{O}(10^{-2})$--$\mathcal{O}(10^{-1})$ rad across a large part of the sensitive frequency band, with some narrowband excursions reported in the observational analyses \cite{Abbott:2021GWTC3}.

To keep the discussion conservative and independent of a specific event, we introduce explicit tolerances $\varepsilon_A$ for amplitude-like effects and $\varepsilon_\phi$ for phase/time--shift effects. Representative numerical estimates will be obtained by considering $\varepsilon\sim 1/\rho$ for a loud event and, alternatively, $\varepsilon\sim 0.1$ as a more conservative tolerance.


\subsubsection{Timelike strain consistency}

In the timelike configuration, the radiation--zone strain is
\begin{equation}
h_{ij}^{(\mathrm{t})\,\mathrm{TT}}(t,r)
=
\frac{2G}{v_t^2 r}
\frac{\mathrm{d}^2 I_{ij}^{\mathrm{TT}}}{\mathrm{d}t^2}
\left(t-\frac{r}{v_t}\right).
\label{strain_timelike_bounds}
\end{equation}
Using
\begin{equation}
v_t\simeq 1+\frac{\xi b_0^2}{2},
\qquad
v_t^{-2}\simeq 1-\xi b_0^2,
\qquad
t-\frac{r}{v_t}
\simeq
t-r+\frac{\xi b_0^2}{2}r,
\end{equation}
one obtains, to first order in $\xi b_0^2$,
\begin{equation}
h_{ij}^{(\mathrm{t})\,\mathrm{TT}}(t,r)
=
\frac{2G}{r}
\left[
\left(1-\xi b_0^2\right)
\frac{\mathrm{d}^2 I_{ij}^{\mathrm{TT}}}{\mathrm{d}t^2}(t_r)
+
\frac{\xi b_0^2}{2}r
\frac{\mathrm{d}^3 I_{ij}^{\mathrm{TT}}}{\mathrm{d}t^3}(t_r)
\right]
+
\mathcal O(\xi^2),
\qquad
t_r=t-r.
\label{timelike_strain_expanded}
\end{equation}
Therefore, the timelike strain contains an amplitude correction
\begin{equation}
\delta A_{\mathrm{t}}=-\xi b_0^2,
\end{equation}
and a phase/time--shift contribution controlled by
\begin{equation}
\delta B_{\mathrm{t}}=\frac{\xi b_0^2}{2}r.
\end{equation}

Requiring the amplitude correction to remain below the effective tolerance $\varepsilon_A$ gives
\begin{equation}
|\delta A_{\mathrm{t}}|\lesssim \varepsilon_A
\qquad\Rightarrow\qquad
|\xi|b_0^2\lesssim \varepsilon_A .
\label{bound_timelike_amp_general}
\end{equation}
For representative choices of the tolerance, this gives
\begin{equation}
|\xi|b_0^2
\lesssim
\begin{cases}
3\times 10^{-2}, & \varepsilon_A\sim 1/\rho\ \text{with}\ \rho\simeq 32,\\[1mm]
10^{-1}, & \varepsilon_A\sim 0.1.
\end{cases}
\label{bound_timelike_amp_numeric}
\end{equation}
This bound is much weaker than the multimessenger propagation bound in Eq.~\eqref{bound_timelike}. It is also partially degenerate with the luminosity distance, inclination, and calibration uncertainties.

For a quasi--monochromatic signal with characteristic angular frequency $\omega\simeq 2\pi f$,
the ratio between the time--shift term and the leading quadrupole contribution is estimated as
\begin{equation}
\left|
\frac{\delta h_{\mathrm{t}}^{(3)}}{h_{\mathrm{t}}^{(2)}}
\right|
\sim
\frac{|\delta B_{\mathrm{t}}|\,\omega}{|1+\delta A_{\mathrm{t}}|}
\simeq
\frac{|\xi|b_0^2}{2}\,r\omega ,
\label{ratio_timelike_phase}
\end{equation}
where the last step holds to leading order in $\xi b_0^2$. Thus, imposing
\begin{equation}
\left|
\frac{\delta h_{\mathrm{t}}^{(3)}}{h_{\mathrm{t}}^{(2)}}
\right|
\lesssim
\varepsilon_\phi
\end{equation}
yields
\begin{equation}
|\xi|b_0^2
\lesssim
\frac{2\varepsilon_\phi}{r\omega}.
\label{bound_timelike_phase_general}
\end{equation}
Using $f\simeq 100~{\rm Hz}$, one obtains
\begin{equation}
|\xi|b_0^2
\lesssim
\begin{cases}
7.7\times 10^{-19}\,\varepsilon_\phi, & r\simeq 40~{\rm Mpc},\\[1mm]
6.2\times 10^{-20}\,\varepsilon_\phi, & r\simeq 500~{\rm Mpc}.
\end{cases}
\label{bound_timelike_phase_numeric}
\end{equation}
For $\varepsilon_\phi\sim 1/\rho\simeq 3\times10^{-2}$, this gives
\begin{equation}
|\xi|b_0^2
\lesssim
2.3\times 10^{-20}
\qquad
(r\simeq 40~{\rm Mpc},\ f\simeq 100~{\rm Hz},\ \rho\simeq 32).
\label{bound_timelike_phase_GW170817}
\end{equation}
Notice that this numerical value should be interpreted as an indicative phase--timing sensitivity, not as a standalone GW--only strain bound, because a frequency--independent time shift can be absorbed into the coalescence time unless an external timing reference is used.


\subsubsection{Spacelike bound from amplitude modulation}

If the quadrupole term is rescaled by a direction-dependent factor, the leading contribution may be written as
\begin{equation}
h_{ij}^{(2)}(t,r)=\frac{2G}{r}\,\Big(1+\delta A(\chi)\Big)\,
\frac{\mathrm{d}^{2}I_{ij}}{\mathrm{d}t^{2}}(t_r),
\qquad t_r=t-r.
\end{equation}
In the present spacelike sector, the explicit radiation--zone waveform shows that the quadrupole
piece carries the prefactor
\begin{equation}
1-\frac{\xi|\mathbf b|^2}{2}\sin^2\chi .
\end{equation}
Matching this expression to the waveform prefactor, one finds
\begin{equation}
1+\delta A(\chi)
=
1-\frac{\xi|\mathbf b|^2}{2}\sin^2\chi
\qquad\Longrightarrow\qquad
\delta A(\chi)
=
-\frac{\xi|\mathbf b|^2}{2}\sin^2\chi .
\label{deltaA_spacelike}
\end{equation}
Therefore,
\begin{equation}
|\delta A(\chi)|
=
\frac{|\xi|\,|\mathbf b|^2}{2}\sin^2\chi
\le
\frac{|\xi|\,|\mathbf b|^2}{2}.
\end{equation}
Within the perturbative regime $|\xi|\,|\mathbf b|^2\ll 1$, the quadrupole amplitude is reduced
for positive $\xi$ and enhanced for negative $\xi$. The modulation vanishes for propagation
parallel to the preferred direction, $\chi=0,\pi$, and is maximal for $\chi=\pi/2$.

Since LVK signals are accurately described by GR templates, a conservative consistency condition is that the fractional amplitude modulation should not exceed an effective tolerance $\varepsilon_A$ over the frequency range carrying most of the matched--filter SNR. In this manner,
\begin{equation}
|\delta A(\chi)|\lesssim \varepsilon_A
\qquad\Rightarrow\qquad
|\xi|\,|\mathbf b|^2
\lesssim
\frac{2\varepsilon_A}{\sin^2\chi} .
\label{bound_amp_general}
\end{equation}
This expression makes explicit the geometric blind spot of the amplitude bound: when
$\chi\simeq 0$ or $\pi$, the modulation disappears and no useful constraint follows from this channel alone. For a representative generic orientation with $\sin^2\chi\sim 1$, one obtains
\begin{equation}
|\xi|\,|\mathbf b|^2
\lesssim
2\varepsilon_A
\sim
\begin{cases}
6\times 10^{-2}, & \varepsilon_A\sim 1/\rho\ \text{with}\ \rho\simeq 32,\\[1mm]
2\times 10^{-1}, & \varepsilon_A\sim 0.1.
\end{cases}
\label{bound_amp_numeric}
\end{equation}
As expected, this amplitude--based constraint is comparatively weak, because an overall
$\mathcal{O}(\xi|\mathbf b|^2)$ rescaling is partially degenerate with extrinsic source parameters, most notably the luminosity distance and the inclination, and also with detector calibration systematics.


\subsubsection{Spacelike phase estimate associated with the $\dddot I_{ij}$ contribution}

The additional higher--derivative term in the spacelike waveform is read directly from the radiation--zone expression,
\begin{equation}
\delta B(\chi)
=
\frac{\xi(b\!\cdot\! r)^2}{2r}
=
\frac{\xi|\mathbf b|^2 r}{2}\cos^2\chi .
\label{deltaB}
\end{equation}
At first order, this contribution can be interpreted as the expansion of the quadrupole term around an angle--dependent shifted retarded time, since
\begin{equation}
\frac{\mathrm{d}^{2}I_{ij}}{\mathrm{d}t^{2}}
\left(t_r+\Delta t_\chi\right)
=
\frac{\mathrm{d}^{2}I_{ij}}{\mathrm{d}t^{2}}(t_r)
+
\Delta t_\chi
\frac{\mathrm{d}^{3}I_{ij}}{\mathrm{d}t^{3}}(t_r)
+
\mathcal O(\Delta t_\chi^2),
\qquad
\Delta t_\chi=
\frac{\xi|\mathbf b|^2 r}{2}\cos^2\chi .
\end{equation}
In other words, the following estimate should be understood as a phase--time--shift criterion rather than as a completely independent GW--only strain constraint.

For a quasi--monochromatic signal at characteristic angular frequency $\omega\simeq 2\pi f$, one expects
\begin{equation}
\left|
\frac{\mathrm{d}^{3}I_{ij}}{\mathrm{d}t^{3}}
\right|
\sim
\omega
\left|
\frac{\mathrm{d}^{2}I_{ij}}{\mathrm{d}t^{2}}
\right|.
\end{equation}
The ratio between the additional term and the quadrupole contribution in Eq.~\eqref{rzrrr} is
therefore estimated as
\begin{equation}
\left|\frac{\delta h^{(3)}}{h^{(2)}}\right|
\sim
\frac{|\delta B(\chi)|\,\omega}{|1+\delta A(\chi)|}
\simeq
|\delta B(\chi)|\,\omega
=
\frac{|\xi|\,|\mathbf b|^2}{2}\,(r\omega)\cos^2\chi ,
\label{ratio_third}
\end{equation}
where the last step holds to leading order in $\xi|\mathbf b|^2$. Imposing
\begin{equation}
\left|\frac{\delta h^{(3)}}{h^{(2)}}\right|\lesssim \varepsilon_\phi
\end{equation}
yields
\begin{equation}
|\xi|\,|\mathbf b|^2
\lesssim
\frac{2\varepsilon_\phi}{r\,\omega\,\cos^2\chi} .
\label{bound_third_general}
\end{equation}
This bound also exhibits a geometric blind spot: for propagation nearly orthogonal to the
preferred direction, $\chi\simeq \pi/2$, the $\dddot I_{ij}$ correction is suppressed and the
constraint weakens. For a representative generic orientation with $\cos^2\chi\sim 1$, we obtain
\begin{equation}
|\xi|\,|\mathbf b|^2
\lesssim
\frac{2\varepsilon_\phi}{r\,\omega}.
\label{bound_third_general_simple}
\end{equation}

Using $f\sim 100~{\rm Hz}$, characteristic of the region of maximal LVK sensitivity, and
representative luminosity distances, we have
\begin{equation}
\frac{2}{r\omega}
\simeq
\begin{cases}
7.7\times 10^{-19}, & r\simeq 40~{\rm Mpc},\\[1mm]
6.2\times 10^{-20}, & r\simeq 500~{\rm Mpc},
\end{cases}
\qquad (f\simeq 100~{\rm Hz}),
\label{two_over_romega}
\end{equation}
so that
\begin{equation}
|\xi|\,|\mathbf b|^2
\lesssim
\begin{cases}
7.7\times 10^{-19}\,\varepsilon_\phi, & r\simeq 40~{\rm Mpc},\\[1mm]
6.2\times 10^{-20}\,\varepsilon_\phi, & r\simeq 500~{\rm Mpc}.
\end{cases}
\label{bound_third_numeric}
\end{equation}
For a loud nearby event such as GW170817, one may take $\varepsilon_\phi\sim 1/\rho\simeq 3\times 10^{-2}$ \cite{Abbott:2017GW170817PRL}, which gives
\begin{equation}
|\xi|\,|\mathbf b|^2
\lesssim
2.3\times 10^{-20}
\qquad
(r\simeq 40~{\rm Mpc},\ f\simeq 100~{\rm Hz},\ \rho\simeq 32),
\label{bound_third_GW170817}
\end{equation}
for a representative generic orientation with $\cos^2\chi\sim 1$. Even adopting the more
conservative tolerance $\varepsilon_\phi\sim 0.1$ yields
\begin{equation}
|\xi|\,|\mathbf b|^2\lesssim 8\times 10^{-20}
\qquad
(r\simeq 40~{\rm Mpc},\ f\simeq 100~{\rm Hz}).
\end{equation}

}


\section{Conclusions}

In this work we investigated the propagation and radiative properties of gravitational waves in the \textit{metric--affine} formulation of the bumblebee gravity scenario, where spontaneous Lorentz symmetry breaking arises from a vector field acquiring a nonvanishing vacuum expectation value. In the Einstein--frame representation of the theory, obtained after integrating out the independent connection, tensor perturbations propagate on the null cone of an effective metric determined by the background vector field. Within the geometric--optics limit of the linearized theory, this structure led to a modified dispersion relation governing the propagation of the graviton modes.

Starting from the relation 
$k^2-\xi\!\left[(b\!\cdot\!k)^2-\frac{1}{2}b^2 k^2\right]=0$,
we examined the consequences of both timelike and spacelike configurations of the background vector $b^{\mu}$. In the timelike sector, the propagation of gravitational waves remained isotropic, but with a modified propagation speed determined by the Lorentz--violating parameter. The corresponding retarded Green function retained the same distributional structure as in general relativity, so that the gravitational waveform preserved the quadrupole form. Nevertheless, the signal was modified through a shift in the retarded time associated with the effective propagation velocity together with a global rescaling of the amplitude.

The spacelike configuration led to a richer radiative structure on the other hand. In this case, the dispersion relation introduced an anisotropic propagation features depending on the orientation between the background vector and the direction of propagation. By constructing the retarded Green function in the Lorentz--violating parameter, we obtained the radiation--zone metric perturbation produced by a compact binary system. The resulting waveform exhibited two characteristic deviations from the standard quadrupole prediction: a direction--dependent modulation of the quadrupole amplitude and an additional radiative term proportional to the third time derivative of the mass quadrupole moment. In the \textit{metric--affine} framework an extra isotropic renormalization of the quadrupole contribution also appeared as a consequence of the pole structure of the modified propagator.

The observational implications of these effects were explored using multimessenger propagation tests and waveform consistency requirements. {The multimessenger comparison between GW170817 and GRB 170817A constrained the propagation speed of the tensor modes, leading to the conservative bounds $|\xi|b_0^2\lesssim 6\times10^{-15}$ in the timelike configuration and $|\xi|b_3^2\lesssim 6\times10^{-15}$ for a spacelike background aligned with the propagation direction. When the asymmetric interval inferred from the multimessenger delay was retained, the corresponding sign--sensitive estimate was written as $-6\times10^{-15}\lesssim \xi b^2\lesssim 1.4\times10^{-15}$, while for a generic spacelike orientation the constrained quantity was the projected combination $|\xi|\,|\mathbf b|^2\cos^2\Psi$.} {At the waveform level, the timelike strain displayed both an amplitude rescaling and an effective phase/time--shift contribution. The amplitude channel yielded only weak consistency limits, of order $|\xi|b_0^2\lesssim \varepsilon_A$, whereas the phase--timing estimate gave $|\xi|b_0^2\lesssim 2\varepsilon_\phi/(r\omega)$ and reached the indicative level of $10^{-20}$ for representative nearby high--SNR events.} {In the spacelike sector, the amplitude modulation produced the geometric bound $|\xi|\,|\mathbf b|^2\lesssim 2\varepsilon_A/\sin^2\chi$, which was comparatively weak and became ineffective when the propagation direction was nearly parallel to the preferred direction. By contrast, the additional $\dddot I_{ij}$ contribution led to the phase--time--shift estimate $|\xi|\,|\mathbf b|^2\lesssim 2\varepsilon_\phi/(r\omega\cos^2\chi)$, which reached the indicative level $\sim 10^{-20}$ for $f\simeq100\,\mathrm{Hz}$ and favorable orientations. }


\section{Acknowledgments}

\hspace{0.5cm}
A. A. Araújo Filho is supported by Conselho Nacional de Desenvolvimento Cient\'{\i}fico e Tecnol\'{o}gico (CNPq) and Fundação de Apoio à Pesquisa do Estado da Paraíba (FAPESQ), project numbers 150223/2025-0 and 1951/2025.
The author also thanks A.~Yu.~Petrov for drawing my attention to the reference on bumblebee gravity in the \textit{metric--affine} framework. The author further thanks P.~J.~Porfírio for valuable discussions and recommendations during the preparation of this work.

\section{Data Availability Statement}

Data Availability Statement: No Data associated in the manuscript


\bibliographystyle{ieeetr}
\bibliography{main}

\end{document}